\documentclass[a4paper,12pt]{article}
\pdfoutput=1
\usepackage{lineno,hyperref}
\usepackage{todonotes}

\usepackage{amsmath}
\usepackage{amssymb}
\usepackage{xcolor}
\hypersetup{
    colorlinks,
    linkcolor={red!50!black},
    citecolor={blue!50!black},
    urlcolor={blue!80!black}
}

\usepackage{authblk}
\usepackage[round]{natbib}

\usepackage{url}
\usepackage{hyperref}

\usepackage{a4wide}

\begin{document}

%%% Maketitle metadata

\title{
		\usefont{OT1}{bch}{b}{n}
		\normalfont \normalsize \textsc{} \\ [25pt]
		\Large Consistency of Regions of Interest as nodes of functional brain networks measured by fMRI\\
}
\author[1,2,*]{Onerva Korhonen}
\author[1]{Heini Saarim\"aki}
\author[1]{Enrico Glerean}
\author[1]{Mikko Sams}
\author[2]{Jari Saram\"aki}

\affil[1]{Department of Neuroscience and Biomedical Engineering, School of Science, Aalto University, Espoo, Finland}
\affil[2]{Department of Computer Science, School of Science, Aalto University, Espoo, Finland}
\affil[*]{Corresponding author: Onerva Korhonen, email: onerva.korhonen@aalto.fi}

\date{}
\maketitle

\begin{abstract}
The functional
network approach, where fMRI BOLD time series are mapped to networks depicting functional relationships between brain areas, 
has opened new insights into the function of the human brain. In this approach, the choice of network nodes is of crucial importance.
One option is to consider fMRI voxels as nodes. This results in a large number of nodes, making network analysis and interpretation of 
results challenging. A common alternative is to use pre-defined clusters of anatomically close voxels, Regions of Interest (ROIs). 
This approach assumes that voxels within ROIs are functionally similar.   
Because these two approaches result in different network structures, it is crucial
to understand what happens to network connectivity when moving from the voxel level to the ROI level. 
We show that the consistency of ROIs, defined as the mean Pearson correlation coefficient between the time series of their voxels, 
varies widely in resting-state experimental data. Therefore the assumption of similar voxel dynamics within each ROI 
does not generally hold. Further, the time series of low-consistency ROIs may be highly correlated, resulting in spurious links in 
ROI-level networks. Based on these results, we recommend that averaging BOLD signals over anatomically defined ROIs should be carefully considered.
\end{abstract}


\section{Introduction} \label{intro}

In the functional brain network approach \citep{sporns2013a, sporns2013b, wig2011}, the brain is depicted
as a collection of nodes and links. Each node represents a brain area that is supposed to be functionally homogeneous, and links represent anatomical or functional 
connections between nodes. 
Studies of structural features of networks constructed from functional magnetic resonance imaging (fMRI)
data have opened new
insights on the dynamics and function of the brain (for reviews, see \citet{papo2014a, bassett2009, sporns2013b}).

However, the reliability of fMRI network analysis
and the factors that affect it have lately become a subject of discussion \citep{aurich2015, shirer2015, shehzad2009, andellini2015, telesford2010, braun2012}.
One of the critical questions is
how to choose what the nodes represent. The anatomical locations and the number of nodes 
have been reported to affect properties of brain networks, such as small-worldness or scale-freeness
\citep{dereus2013, wang2009, zalesky2010, sporns2013a, sporns2013b}.

It would be natural to consider single neurons as network nodes, linked by synaptic connections. However, 
this micro-level approach is not an option for studying the whole human
brain, because of 
the excessive number of neurons and 
the lack of resolution of imaging methods \citep{dereus2013}.
Thus, network studies of the brain are currently limited to
the mesoscopic and macroscopic levels. For fMRI, these are the level of imaging
voxels and the level of Regions of Interest (ROIs) that are collections of voxels defined on the basis of
\emph{e.g.} anatomical landmarks.

In the voxel-level approach, the nodes are voxels, cubical volume elements with  2--8  
mm edges that form a regular grid covering the brain.
The BOLD signal associated with each voxel is directly given by the fMRI measurement \citep{stanley2013}.
In the ROI-level approach, the nodes are ROIs that comprise tens to hundreds of voxels. A ROI's signal 
is typically computed by averaging the BOLD signals of its voxels \citep{stanley2013}.
ROIs are usually defined using  
an anatomical atlas, based on  
structural MR images or histological investigations (see, \emph{e.g.}, \citet{stanley2013}).  
Data-driven methods that cluster voxels based
on resting-state functional connectivity (\emph{e.g.} \citet{nelson2010, power2011, craddock2012, shen2013}, for a review see
\citet{sporns2013b, wig2011}), ICA, and dual regression \citep{beckmann2009}, or combination of anatomical, functional, and connectivity data
\citep{glasser2016multi, fan2016human} have also been suggested for defining ROIs. 
However, despite promising results, these methods are used only infrequently.

The main benefits of the ROI approach are 
increased signal-to-noise ratio (SNR) and decreased computational cost. 
Further, one can expect that cognitive functions cover brain areas larger
than single voxels \citep{wig2011, shen2013}. Therefore, the ROI approach may characterize true brain activity better than voxels with functionally arbitrary boundaries. 
However, areas related to cognitive functions may not necessarily match with the anatomical boundaries that define ROIs: 
the same function may be distributed across multiple
anatomical areas, or one anatomical area may contain several functionally distinct subareas \citep{stanley2013, papo2014b}.
Thus, the main disadvantage of the ROI approach is the possible loss of information that results from averaging signals of voxels that represent  different functions for producing the ROI signal \citep{stanley2013}.

On the other hand, if the functional network is constructed with voxels as nodes, problems arise from the numbers of nodes and links that represent 
correlations between voxel time series. Because of computational limitations, one has to heavy-handedly threshold
correlation matrices to reduce the number of links. This 
results in sparse networks where much information has been discarded. 

In both approaches, some information is deliberately discarded. But do they retain \emph{similar} information? There is some evidence that moving from the voxel level
to the ROI level changes network properties. \citet{hayasaka2010} found that  ROI-level networks are less robust against  
fragmentation than voxel-level networks at low network densities. Further, ROI-level networks showed less small-world properties, and had different degree distributions
and less stable hubs.
\citet{tohka2012} reported similar results in
structural brain networks that were based on the thickness of cortical gray matter. Therefore, one may ask if voxel-level and ROI-level networks provide comparable views on the underlying brain function.

In this article, we ask what happens to network connectivity when moving from the voxel-level network to the ROI level, where in both cases links represent zero-lag correlations. 
In particular, we focus on whether voxels of a ROI 
display coherent dynamics, as they should if the ROIs match with underlying functional areas. 
To this end, 
we introduce the concept of ROI \emph{consistency} that quantifies the similarity of the signals of voxels that comprise the ROI, defined as their mean Pearson correlation 
coefficient. With the help of resting-state fMRI data of 13 subjects measured in-house as well as 28 subjects from the ABIDE initiative, we show that consistency varies widely across ROIs. Therefore voxel dynamics within ROIs are not always
coherent. We then show that this variation is reflected in network properties: signals of voxels in non-consistent ROIs are not correlated, and less consistent ROIs are 
less central as nodes of the functional brain network.

\section{Results} \label{results}

\subsection{Voxel time series within ROIs are on average correlated, but not uniformly} \label{results:within_roi_correlations}

We begin by considering the functional homogeneity of voxels within ROIs -- in the ROI approach, it is assumed that voxels within a ROI are functionally more similar and thereby have 
more strongly correlated time series than voxels in different ROIs. For defining ROIs, we use three different atlases: the anatomical HarvardOxford (HO) and Automated Anatomical Labeling
(AAL) atlases as well as the Brainnetome atlas that is based on structural and functional connectivity. 

For testing the homogeneity assumption, 
we calculated the distribution of Pearson correlation coefficients between the time series of pairs of voxels that are in the same ROI. 
This distribution was calculated across all ROIs and therefore covered the whole cerebral cortex and subcortical structures. In HO and AAL, also cerebellum was included; the Brainnetome
atlas does not contain cerebellar ROIs. As a reference, 
we computed a similar distribution for pairs of voxels that reside in different ROIs. 

\begin{figure}[]
  \begin{center}
      \includegraphics[width=0.5\linewidth]{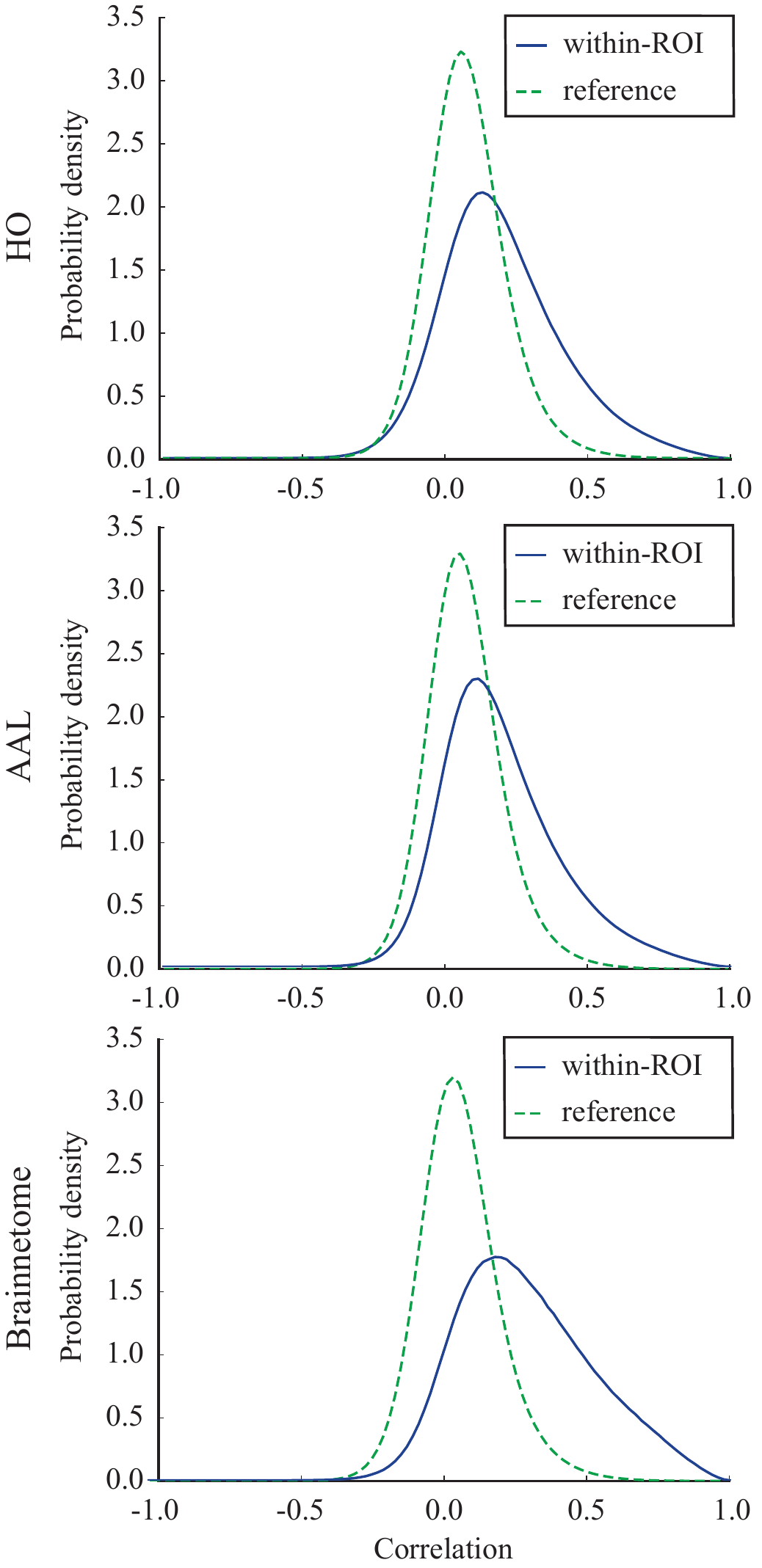} 
      \caption{Voxel time series are on average more strongly correlated within ROIs than between ROIs but the correlation is not perfect. Distributions of Pearson correlation 
coefficients between voxel time series within (blue) and between (green, reference) ROIs in HO, AAL, and Brainnetome. Although the within-ROI correlation distribution is more skewed to the right, 
the distributions largely overlap and small and negative correlations exist also within ROIs. Correlation distributions are calculated across all ROIs from the pooled data of 13 subjects.}
      \label{within_roi_correlations}
 \end{center}
\end{figure}

Correlations are on average stronger within ROIs than in the reference (\textit{i.e.} voxels
in different ROIs) in all investigated parcellations: their distribution has a 
higher mean (HO: $r=0.20$ vs $r=0.073$, Student's $t=49.81$, $p\ll 10^{-5}$; AAL: $r=0.19$ vs $r=0.067$, $t=50.00$, $p\ll 10^{-5}$; Brainnetome: $r=0.29$ vs $r=0.073$, $t=83.05$, 
$p\ll 10^{-5}$)\footnote{The $p$-value has been calculated using a 
permutation-based two-tailed $t$-test \citep{glerean2016reorganization}.}
and is more strongly right-skewed than the approximately normal reference distribution (Fig~\ref{within_roi_correlations}). However, there is a large overlap between the two distributions,
and a significant number of small and negative correlations is present in both distributions. Thus, the functional uniformity of ROIs is far from perfect. 

One might expect the connectivity-based Brainnetome parcellation to show stronger functional homogeneity of ROIs than the anatomical HO and AAL. Surprisingly, this is not the case:
although the mean of the within-ROI correlation distribution is sligthly higher in Brainnetome than in other parcellations, the overlap of the within-ROI and reference correlation distributions
is equally large in all parcellations.

In order to investigate, to which extend our results generalize to other datasets, we repeated all analysis of this article on a second, independet dataset, the ABIDE data \citep{di2014autism}. The results obtained
using the ABIDE data were highly similar to those obtained using the in-house dataset and presented in this Results section. A detailed description of the results obtained using the ABIDE
dataset can be found in Supplementary Results.

Spatial smoothing, a commonly used method in the preprocessing of fMRI data, makes the time series of neighbouring voxels more similar, 
and is therefore expected to affect the results presented above. We repeated all analysis of this article for data that were smoothed with three Gaussian kernels with different
full widths at half maximum. Although spatial smoothing increased all correlations between voxel time series, it has no qualitative effects on the results; for example correlation distributions are
equally broad after spatial smoothing as without it (for a detailed description of the effects of spatial smoothing, see Supplementary Results).

\subsection{The consistency of ROIs varies across the brain} \label{results:consistency_varies}

\begin{figure}[]
  \begin{center}
      \includegraphics[width=0.8\linewidth]{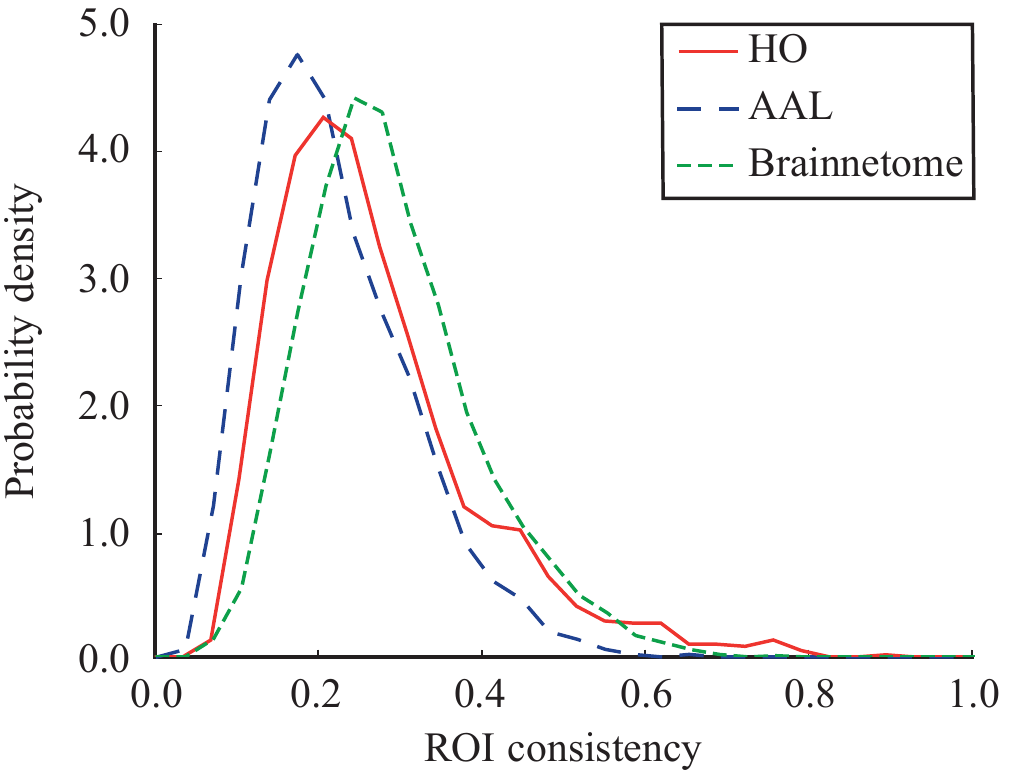}  
      \caption{Consistency of ROIs is widely distributed. The consistency distributions has been calculated
from the pooled data of 13 subjects.}
      \label{consistency_dist}
 \end{center}
\end{figure}

The consistency of a ROI ($\phi$), defined as the mean Pearson correlation coefficient of the time series of voxels belonging to the ROI, is a simple measure of the ROI's functional cohesion. 
In order to evaluate the variation of consistency between ROIs, we calculated consistency distributions for each parcellation from the pooled data of all 13 subjects (Fig~\ref{consistency_dist}). 
Although the maximum observed ROI consistency is high (HO: $\phi_\mathrm{max} = 0.89$; AAL: $\phi_\mathrm{max} = 0.66$; Brainnetome: $\phi_\mathrm{max} = 0.75$), the consistency 
distribution is broad, peaking at low consistency values (HO: $\phi=0.22$; AAL: $\phi=0.18$; Brainnetome: $\phi=0.25$).
This indicates that in all the three atlases, a significant number of ROIs contain collections of voxels that are not functionally uniform in the resting state. 

\begin{figure*}[]
  \begin{center}
      \includegraphics[width=0.9\linewidth]{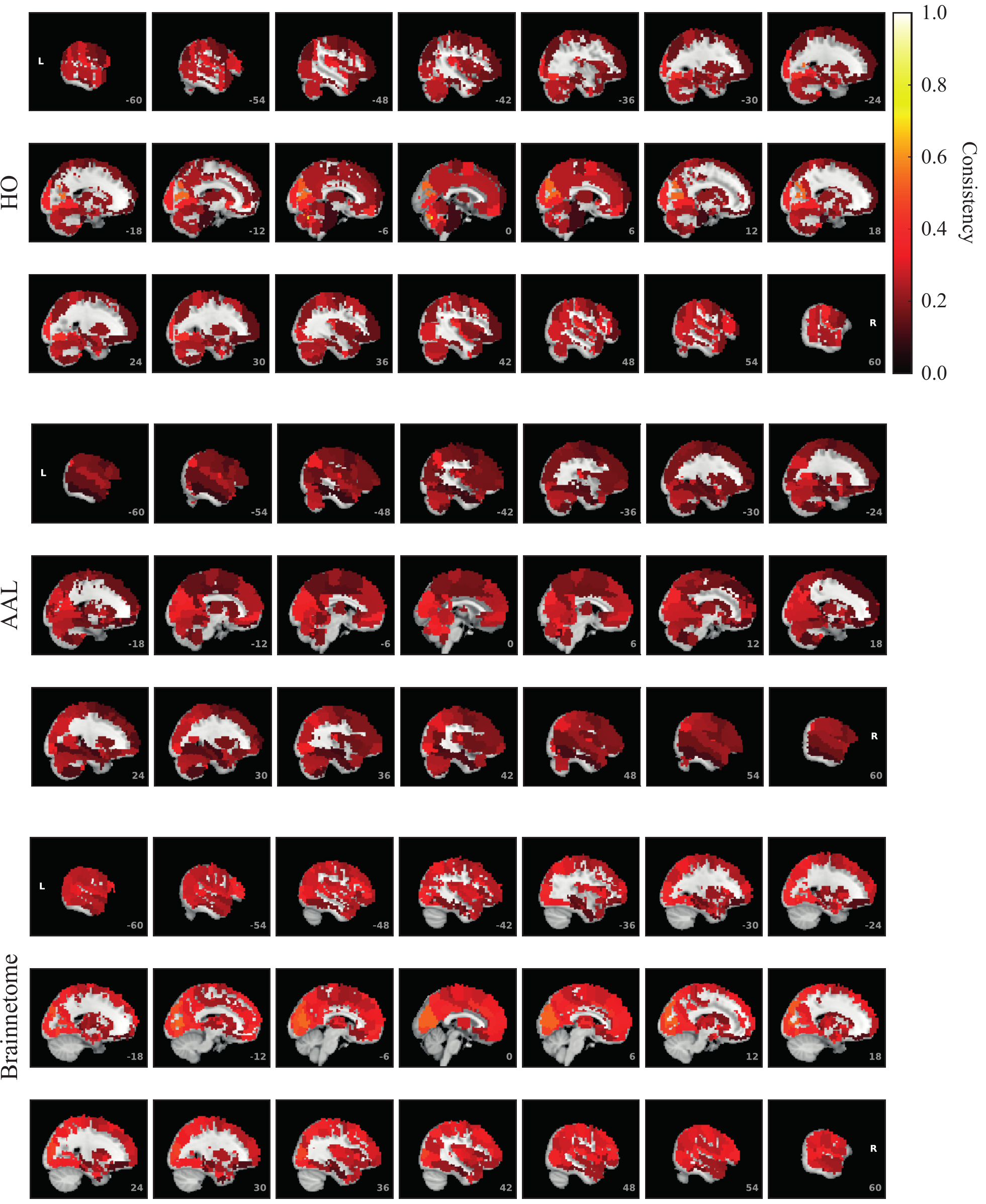}  
      \caption{ROI consistency is not anatomically uniform. We have visualized the mean consistency across 13 subjects on a template brain. ROIs colored with yellow and orange are those
      with highest consistency, while low-consistency ROIs are colored with black. See the main text for the details of the identity of these ROIs. Areas colored with grey-white scale 
      (white matter and, in the case of Brainnetome, cerebellum) are not included in this study.}
      \label{consistency_on_brain_surface}
 \end{center}
\end{figure*}

To investigate in more detail how consistency is distributed among ROIs, we calculated the mean consistency for each ROI across subjects. Visualization
of the mean consistency on a brain template (Fig~\ref{consistency_on_brain_surface}) demonstrates what the broad consistency distribution means in practice: the most consistent 
ROIs have twice the consistency of the least consistent ones. 

The identity of the most consistent ROIs varies across subjects only slightly. In HO, the most coherent ROIs include the
left and right supracalcarine, cuneal, and intracalcarine cortices as well as the cerebellar areas Vermis Crus II and Vermis VIIb. In the case of the supracalcarine cortices and 
the cerebellar areas the high consistency is probably explained by the size of the ROIs, since all these areas are small (containing 10 voxels or less).

In AAL, the most consistent ROIs included left and right cuneal cortex and right Heschl's gyrus as well as the small cerebellar ROIs Vermis\_1\_2 and Vermis\_10. In Brainnetome,
the most consistent ROIs included several subsections of the left and right cuneal cortex and right medial superior occipital gyrus.

For the least consistent ROIs, intersubject variation is larger. In HO, the least consistent ROIS include the right frontal orbital cortex, right and left frontal pole, right precentral gyrus, 
and superior division of left lateral occipital cortex. In AAL, among the least consistent ROIs were left and right hippocampus, left and right inferior temporal gyrus, and left frontal
superior orbital gyrus. In Brainnetome, the lest consistent ROIs included left and right orbital gyrus (6\_3), left and right fusiform gurys, and left and right hippocampus.

ROIs of different parcellations do not perfectly match with each other. However, it is interesting to see that same ROIs, in particular cuneal cortices, are among the most consistent
ROIs in all three parcellations. For sizes, mean consistency values, and consistency ranks of all ROIs, see Supplementary Table.

In terms of mean consistency, the differences between the three parcellations investigated are small. Taking into account the individual variation of structure and function of the brain,
one might expect to see more variation in consistency between subjects in Brainnetome than in HO and AAL. In order to test this hypothesis, we calculated the standard deviation of 
consistency across subjects for each ROI. However, we found no significant difference between the standard deviations of consistency in different parcellations.

\begin{figure}[]
  \begin{center}
      \includegraphics[width=0.5\linewidth]{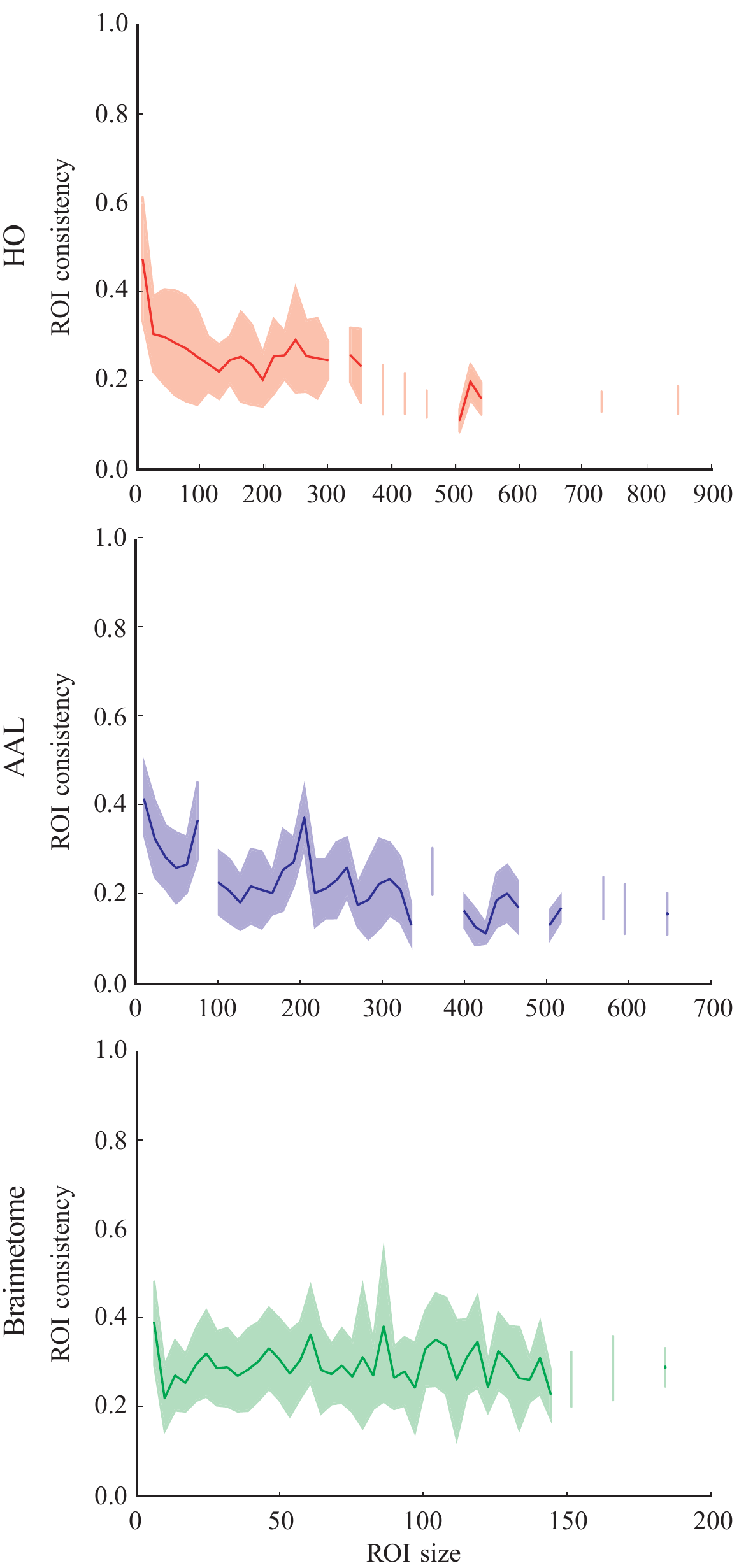}    
      \caption{The size dependence of the average ROI consistency (line) and its variation (shadowed area) show that, in HO and AAL, small ROIs have on average higher consistency.  
      This dependency is less visible for ROIs larger than one hundred voxels. However, the variation of consistency is large for all ROI sizes. Therefore, the size of a ROI is not the 
      only factor that affects its consistency. In Brainnetome, ROI size and consistency are not correlated. The consistency of each ROI has been pooled across 13 subjects, binned on the basis of ROI size, and then bin-averaged.}
      \label{roi_size_vs_consistency}
 \end{center}
\end{figure}

In HO and AAL parcellations, ROI consistency is partially explained by its size, \emph{i.e.} the number of voxels within the ROI (Fig~\ref{roi_size_vs_consistency}).
Consistency is highest for small ROIs and decreases with increasing ROI size (HO: Pearson correlation coefficient $r=-0.35$, $p\ll 10^{-5}$; AAL: $r=-0.38, p\ll 10^{-5}$). 
This decrease saturates for ROIs larger than  a few hundred voxels. However, for ROIs of any size, there is a lot of variation around the mean. In Brainnetome, in contrast, ROI consistency
and size do not correlate ($r=0.017$, $p=0.35$).

We investigated also if head motion explains some part of the individual differences in ROI consistency. However, we did not find significant correlation between ROI consistency and subjects' 
head motion, measured in terms of framewise displacement.

\subsection{Low ROI consistency predicts low voxel-level correlations between pairs of ROIs} \label{results:voxel_level_correlations}

One of the  aims of this paper is to understand how the consistency of ROIs relates to their network properties. We begin by looking at the voxel level, and 
define the
voxel-level correlation between two ROIs as the mean Pearson correlation coefficient between time series of voxels within the ROIs (see Eq~\ref{eq:voxel-level}). 
We expect that the voxel-level correlation is strong only if the ROIs are consistent enough. This is motivated as follows: 
a strong voxel-level correlation means that many pairs of voxels in both ROIs must have similar time series. This means that in each ROI, there must be a lot of similarly 
behaving voxels, resulting in high consistency. As the limiting case, $ \left \langle C\left(I,J\right)\right \rangle_\mathrm{vox}=1$ if and only if the time series of all 
voxels in both ROIs are equal ($x_i=x_j \forall i\in I,j\in J$ and $x_i=x_{i'} \forall i,i'\in I$). 
In order to test this hypothesis, we investigate voxel-level correlations between two ROIs as a function of the mean consistency of the ROI pair. 

As we hypothesized, our data indicate that high mean consistency is a 
prerequisite for a strong voxel-level correlation between two ROIs (Fig~\ref{consistency_vs_correlation}, left). However, it is clear that there is a lot of variation: 
 voxel-level correlation correlates only moderately with ROI consistency (HO: Pearson correlation $r=0.29$, $p\ll 10^{-5}$; AAL: $r=0.17$, $p\ll 10^{-5}$; Brainnetome:
 $r=0.28$, $p\ll 10^{-5}$).
Despite this variation, voxel-level correlations are never strong between low-consistency ROIs, and all data points are 
 located below the identity line ($\left \langle C\left(I,J\right)\right \rangle_\mathrm{vox} = \phi$). 

\subsection{Low-consistency ROIs may have high ROI-level correlations} \label{results:roi_level_correlations}

\begin{figure*}[]
  \begin{center}
      \includegraphics[width=0.7\linewidth]{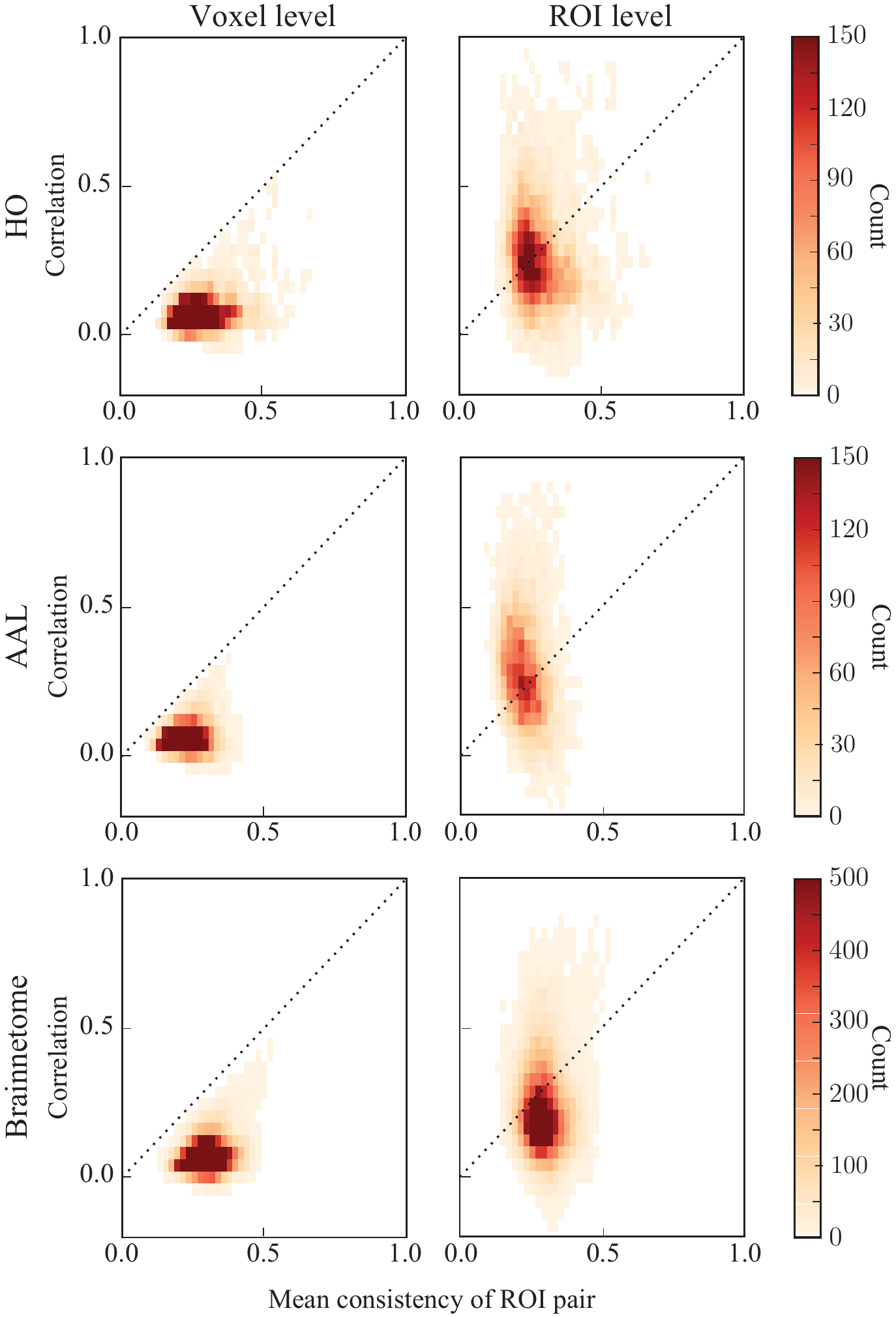}      
      \caption{Averaging of voxel signals may induce spuriously strong ROI-level correlations between ROIs that have low consistency. Left: The relationship between voxel-level correlation and the mean ROI consistency for each ROI pair. 
Voxel-level correlation correlates with consistency, and all data points are located below the identity line. Right: the same relationship for ROI-level correlations and mean ROI consistency.  ROI-level correlations are not correlated with consistency, and strong ROI-level correlations also exist between ROIs that have low consistency and weak voxel-level correlation. In order to produce the heatmaps, the consistency and voxel and ROI-level correlations have been averaged across 13 subjects.}
      \label{consistency_vs_correlation}
 \end{center}
\end{figure*}

Next, we investigated how ROI consistency relates to correlations at the ROI level. To this end, for pairs of ROIs, we calculated the ROI-level correlations (see Eq~\ref{eq:roi-level}) 
as a function of their mean consistency.
In HO and Brainnetome, ROI-level correlations depend on consistency clearly less than the voxel-level correlations (Fig~\ref{consistency_vs_correlation}, right; HO: Pearson correlation 
$r=-0.14$, $p\ll 10^{-5}$; Brainnetome: $r=-0.00057$, $p=0.92$), whereas in AAL ROI consistency and ROI-level correlation are negatively correlated ($r=-0.24$, $p\ll10^{-5}$).

As seen in Fig~\ref{consistency_vs_correlation} (right), there are pairs of low-consistency ROIs that nevertheless display strong ROI-level correlations. 
From the functional point of view, this situation is not straightforward to interpret: if strong ROI-level correlations are taken as a sign of a strong functional 
relationship, how can this relationship be real if the ROIs themselves are nonuniform and lack consistency?

As one possible reason for these spurious-looking correlations,
let us assume that signals of voxels in ROIs $I$ and $J$ share 
a common component, but their pairwise correlations are  weak  because of components unique to the signal of each voxel. In this case, the consistency of both ROIs is low.
Then, the time series of the ROIs are obtained as averages over the time series of their voxels (see Eq~\ref{eq:roiseries}).
This averaging amplifies signal components shared by all voxels within the ROI and suppresses components unique to each voxel.
Therefore, the time series of ROIs $I$ and $J$ consist mostly of the shared signal component and are strongly correlated. 

The signal component shared by each voxel in the ROIs $I$ and $J$ may be either noise or true signal. Similarly, the suppressed signals components may be either true
signal components unique to single voxels or independent noise. Based on the results visualized in Fig~\ref{consistency_vs_correlation}, it is not possible to say for sure
if the links between low-consistency ROIs are spurious or if they are true correlations that have become visible when noisy signal components have been suppressed. However,
this result suggest that extra care is needed when interpreting the ROI-level correlations, since any 
artifactual signal shared between voxels can induce spurious ROI-level correlations between ROIs that have low consistency.

\begin{figure}[h!]
  \begin{center}
      \includegraphics[width=0.6\linewidth]{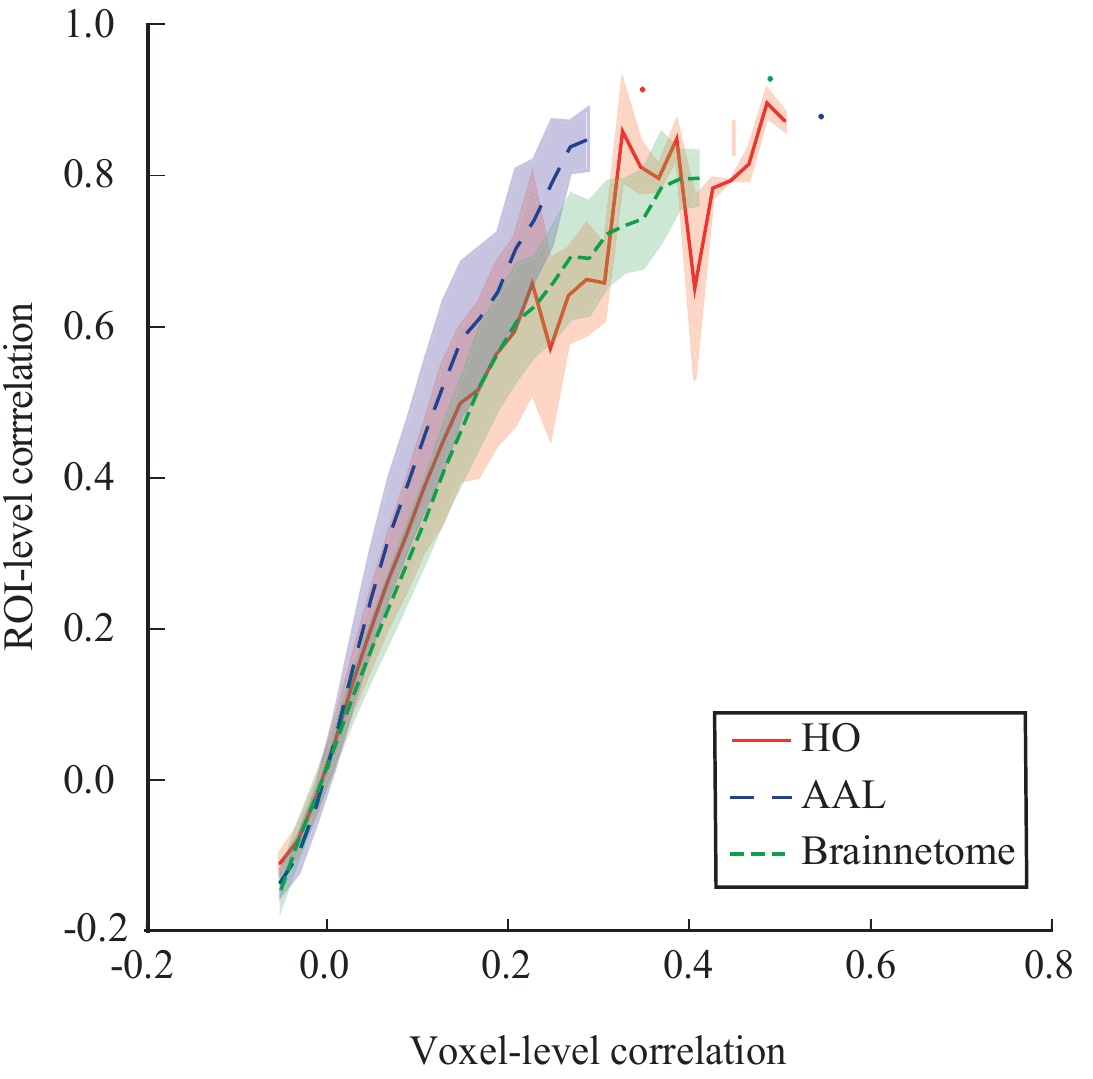}  
      \caption{ROI-level correlations are stronger than the voxel-level correlations between the same ROIs. This is because averaging the voxel signals in order to obtain the ROI time series
      suppresses individual signal components of each voxel and amplifies components shared among voxels (for details and discussion, see the main text). 
 Averaging and binning as in Fig.~\ref{roi_size_vs_consistency}.}
    \label{roi_vs_voxel_level_corr}
 \end{center}
\end{figure}

For a concrete example of the effects of amplification and suppression of voxel-level
signal components when moving to the ROI level, we investigated ROI-level correlations as a function of voxel-level correlations (Fig~\ref{roi_vs_voxel_level_corr}).
We observed that in all investigated parcellations, on average ROI-level correlations increase faster than the increasing voxel-level correlations: pairs of ROIs that correlate only moderately at the voxel level can 
appear highly correlated at the ROI level. 
For instance, a ROI pair that has a voxel-level correlation of $r=0.2$ can have a ROI-level correlation as high as $r=0.6$.
This is, of course, a direct result of the averaging of voxel signals in order to obtain the ROI time series. 

Further, the relationship between ROI-level and voxel-level correlations is nonlinear and saturates already at rather low voxel-level correlation values. Therefore, pairs of ROIs that appear to have similar correlations at the ROI level may in fact clearly differ at the voxel level. 

\subsection{Consistency of a ROI predicts its network properties} \label{results:nproperties}

\begin{figure*}[h!]
  \begin{center}
      \includegraphics[width=0.8\linewidth]{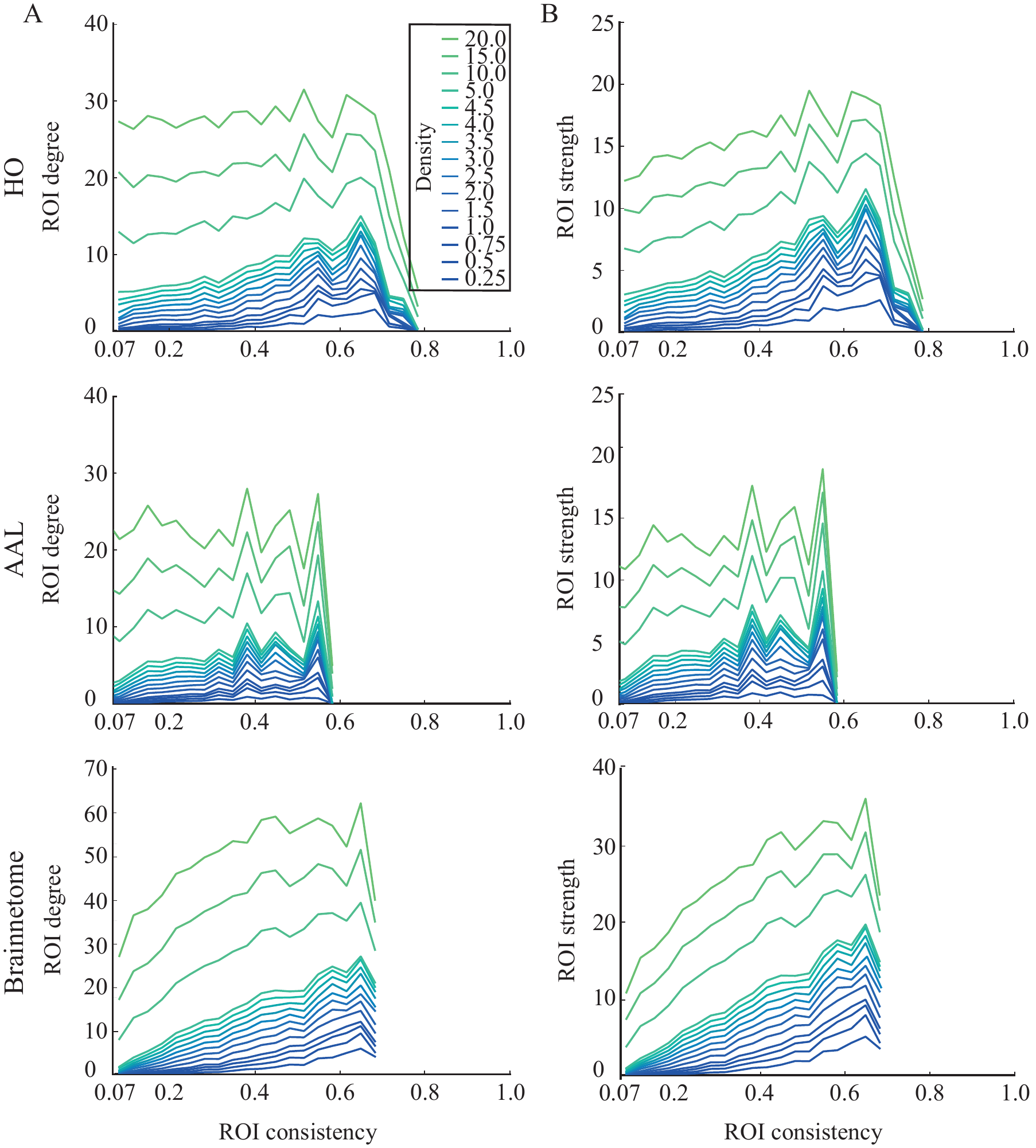}  
      \caption{Degree and strength of ROIs increase with consistency. For degree (A), this behavior is mostly visible only at low and intermediate network densities, whereas strength (B) and consistency correlate even for high densities. Averaging and binning as in Fig.~\ref{roi_size_vs_consistency}.}
    \label{consistency_vs_roi_properties}
 \end{center}
\end{figure*}

The typical way of obtaining functional brain networks from correlation matrices is to threshold them so that only the strongest links are retained. We next 
thresholded the ROI-level networks, in order to investigate if the consistency of a ROI affects its network properties. For the thresholded networks, we calculated
the mean degree ($k$)
and strength ($s$) of ROIs across subjects. We observed that both the degree and the strength 
(Fig~\ref{consistency_vs_roi_properties})
increase with increasing consistency in networks thresholded to low and intermediate densities. For example, at $d=0.25\%$, the Pearson correlation coefficient for 
degree is in HO $r=0.39$, $p\ll 10^{-5}$, in AAL $r=0.27$, $p\ll10^{-5}$, and in Brainnetome $r=0.53$, $p\ll10^{-5}$,
and the Pearson correlation coefficient for strength 
is in HO $r=0.39$, $p\ll 10^{-5}$, in AAL $r=0.28$, $p\ll
10^{-5}$, and in Brainnetome $r=0.53$, $p\ll10^{-5}$.

Above, we showed that  ROI-level correlations between pairs of ROIs appear rather independent of their consistencies. This may appear at odds with the degree and strength increasing 
with consistency; however, the independence of ROI-level correlations on consistency shows in full correlation matrices. When the correlation matrices are thresholded to retain only 
the strongest links, it becomes evident that these strongest links often connect to high-consistency ROIs.

In networks thresholded to higher densities, degree and strength are more independent of consistency. In HO and AAL, at the network density of $d=20\%$, no correlation is present between neither ROI 
consistency and degree (HO: Pearson correlation coefficient $r=-0.055$, $p=0.81$; AAL: $r=-0.039$, $p=0.13$) nor ROI consistency and strength (HO: $r=0.098$, $p=3.2*10^{-5}$; AAL: $r=0.049$, $p=0.057$).
In Brainnetome, a weak correlation between consistency and degree remains at density of $d=20$ ($r=0.23$, $p\ll10^{-5}$), and consistency and strength are correlated both at density
$d=20$ ($r=0.036$, $p\ll10^{-5}$) and in the full network ($r=0.19$, $p\ll10^{-5}$).

\section{Discussion} \label{discussion}

\subsection{Variation in consistency of ROIs points towards differences in their functional homogeneity} \label{discussion:inhomogeneities}

The common approach of averaging over voxel time series in order to obtain representative time series for entire ROIs does not account for possible variations and inhomogeneities within 
the ROI. Therefore, to measure the level of such inhomogeneities, we introduced the concept of ROI consistency ($\phi$) that is defined as the mean Pearson correlation coefficient between the time 
series of voxels in the ROI. This measure is simple to calculate and easy to interpret. If a ROI has high consistency, one can assume that the averaged time series accurately represent 
the underlying voxel dynamics. On the other hand, low consistency can be seen as indicative of functional inhomogeneity in the ROI.

Earlier, measures similar to ROI consistency have been used to quantify multiple aspects of the human brain function. For example, \citet{li2002alzheimer} have suggested that
the consistency of the hippocampus could be used as an early-stage biomarker for Alzheimer's disease.

We investigated ROI consistency in three different atlases: the anatomical HarvardOxford (HO) and Automated Anatomical Labeling (AAL) atlases and Brainnetome atlas that is based on structural
and functional connectivity. We found that distribution of consistencies across ROIs is broad in all these atlases. The most consistent ROIs are twice as consistent as the least consistent ones. This is in line with earlier observations \citep{baumgartner1999, craddock2012}:
voxel time series within an activated brain area are not always strongly correlated. Further, the anatomical distribution of consistency is not uniform. In contrast to what one might
expect, we found no significant difference between the Brainnetome atlas and the anatomical atlases in terms of mean consistency of ROIs.

\subsection{What does consistency tell about the properties and functional role of a ROI?} \label{discussion:consistency}

It is tempting to interpret the consistency as a measure of the \emph{goodness} of a ROI: ROIs with high consistency are well-defined, whereas low consistency indicates inaccuracies in the ROI
definition. Indeed, a measure similar to mean consistency across ROIs has been used to quantify the overall accuracy of a parcellation (\citet{craddock2012, thirion2006, stanley2013};
see also \citet{gordon2014generation} for a different definition of ROI homogeneity based on network connectivity). However, this interpretation, although useful, is probably oversimplified. 

The non-uniform anatomical distribution of consistency opens the question of its  dependency on time and task. Does
the consistency of a ROI vary in  
time and how is it determined at a given moment? One plausible hypothesis is that active brain 
areas have higher consistency than currently inactive areas. This would also explain 
 the dependency between consistency and degree of ROI at low-to-intermediate network densities. More active ROIs that should have higher consistency also have stronger momentary functional connections to other active ROIs and thus have a 
higher degree. 

The concept of Regional Homogeneity (ReHo) (\citet{zang2004}, see also \citet{jiang2016regional} for a review on how ReHo has been used) is based on an assumption similar to the above hypothesis on the increased consistency of active brain regions.
In the ReHo framework, active voxels are identified by their increased similarity to their neighbours. This similarity is calculated in a cubic neighbourhood that includes the voxel itself and its neighbours
in terms of faces (7 voxels), faces and edges (19 voxels), or faces, edges, and corners (27 voxels). Despite minor technical differences -- in ReHo, Kendall's coefficient
of concordance is used to quantify the similarity of voxel time series --, the similarity of ReHo and ROI consistency is obvious: if we define a ROI that covers the ReHo neighbourhood of a voxel, 
the consistency of this ROI is proportional to the ReHo value of the centroid voxel. 

ReHo has been used to quantify the changes in local connectivity not only in single voxels but also in brain areas.
\citet{jiang2015toward} calculated ReHo using a two-dimensional reconstruction of the cortex, and averaged the ReHo values
across voxels within functionally defined subareas in the ventral visual system, the prefrontal cortex, and the posteromedial cortex. The differences of the mean ReHo between the subareas, obtained
during resting state, reflected the different roles of these subareas in hierarchical information processing. 

Despite the obvious similarities between consistency and ReHo, the purpose of these two measures is different. ReHo is a voxel-level measure that is always calculated across the immediate
neighbourhood of a voxel. Originally, ReHo has been designed to localize activity to 
a handful of voxels that may be spatially separated \citep{zang2004}, and it can be used also to construct functional brain networks at the level of voxels \citep{jiang2015toward}. To the contrary, 
consistency is a property of a collection of voxels (ROI) and can be used to quantify how well the average signal of these voxels represent the original voxel-level dynamics. Therefore,
the usage of consistency is by definition limited to cases where activated brain areas -- and nodes of functional brain networks -- are assumed to be larger than single voxels. As consistency
is by definition a ROI-level measure, it is natural that it relates to properties of ROI-level brain networks, for example node centrality measures and strength of links.

Bearing these differences between ReHo and consistency in mind, it is logical that
a ROI with low consistency can still contain voxels that have high ReHo; this is the case if the ROI definition is not accurate and the ROI 
contains subareas of internally correlated voxels. If the boundary between these subareas is sharp enough, even the mean ReHo of this low-consistency ROI can be high.
Similarly, a ROI with high consistency can contain low-ReHo voxels: one can expect to find these at least at the boundaries of the ROI where the ReHo neighbourhood contains
voxels from different ROIs. 

Task-induced changes in the consistency of a ROI could also tell about its functional role, similarly as the mean ReHo in the study of \citet{jiang2015toward}. On one hand, if a ROI consists 
of functionally different subareas, some tasks may require synchronous activation of them all, yielding high consistency. On the other hand, in some conditions these subareas may activate at different times, which 
results in lower consistency. Therefore, the application of the concept of consistency beyond single ROIs, \emph{e.g.} to resting state or task-related networks,
may reveal dynamics of co-activation and separation of different subnetworks. 

High values of test-retest reliability have been reported for several measures of local connectivity, in particular for ReHo \citep{zuo2013toward, zuo2014test}. In the present study, we 
do not investigate the test-retest reliability of consistency: when consistency is used to evaluate the level
of within-ROI homogeneity, low consistency values measured in one resting-state session can be viewed as a sufficient counter-example
to the assumption of homogeneous voxel dynamics within a ROI. However, when task-dependency of consistency is investigated and, in particular, if one assumes that high consistency
is indicative of increased activity, it is important to address the question of the test-retest reliability of the measure. In other words, before investigating the changes
of consistency between rest and task, it might be good to ensure that consistency is stable between different resting-state sessions of the same subject. 
This investigation should address also the possible effects of preprocessing on the test-retest
reliability, similarly as \citet{zuo2013toward} have been done for ReHo.

\subsection{The consistency of ROIs is affected by their shape and size and the level of spatial smoothing} \label{discussion:roi_size}

 There are multiple factors that can explain the consistency of a ROI. First, even if one assumes that perfectly consistent and static ROIs exist, atlases 
 have typically been constructed at group level and do not perfectly match any individuals. Because of this, a ROI may overlap with many true functional subareas that are 
 internally coherent, and following this, small ROIs are less likely to consist of several subareas. Similar results have been obtained also by \citet{gordon2014generation} using a network-level homogeneity measure. Furthermore, adjacent voxels are known to be more strongly 
 correlated~\citep{alexander2012, salvador2005}. Therefore, small ROIs should be more consistent. Additionally, the shape of a ROI may affects its consistency as well: 
 small, spherical ROIs should be most consistent. 
 
Interestingly, we found that consistency depends on ROI size in the anatomical atlases but not in Brainnetome. In Brainnetome, ROIs are on average smaller than in HO and AAL
and the variation of ROI size is also smaller, which may explain this difference between atlases \citep{gordon2014generation}. However, it is also possible that in the connectivity-based Brainnetome atlas,
ROI consistency is less explained by size and more by other features such as the functionality of the ROI.
 
Spatial smoothing is commonly used in the preprocessing of fMRI data. We found that spatial smoothing increases
the consistency of practically every ROI. However, smoothing does not make the consistency distribution significantly more narrow. Even when smoothing is applied, the consistency values of the 
least consistent ROIs are only half of those of the most consistent ROIs. 

From the viewpoint of voxel signals, the effects of spatial 
smoothing are qualitatively similar to those of averaging voxel signals in order to obtain ROI time series: any shared signal components are amplified whereas individual components are suppressed. Therefore, spatial smoothing
increases the correlations between all voxel time series. This increase is strongest for voxels that are less than a FWHM apart, which is typically 
the case for voxels within the same ROI. In conclusion, one may argue that spatial smoothing is redundant when constructing ROI-level networks: noise suppression is already taken care of when the voxel-level signals are averaged to ROI-level signals, and smoothing basically only results in a baseline shift in ROI consistencies.

\subsection{Averaging of voxel signals may produce spuriously high ROI-level correlations} \label{discussion:spurious_links}

We demonstrated that there are pairs of low-consistency ROIs that have strong ROI-level correlations; in general, ROI-level correlations are scattered and do not appear to depend on ROI consistency. To the contrary, the mean consistency of a ROI pair determines an upper limit for their voxel-level correlation. 
The existence of spuriously high ROI-level correlations is explained by the amplification of shared voxel-level signal components when the voxel signals are averaged in order to obtain the ROI time series. At the same time, averaging suppresses individual components. 

Averaging related to moving from the voxel level to the ROI level can be seen from two different angles. First, if we assume that ROIs represent functionally uniform groups of voxels that have
an underlying common signal component, then the individual voxel-level signal components can be considered as
 noise. In this case, averaging acts as a filter that suppresses noisy signal components and increases the signal-to-noise ratio (SNR). On the other hand, individual voxel-level signal
components can also reflect true underlying neural activity, and ROI-wide shared components can be caused by 
physiological noise or external noise. Should this be the case, averaging of the voxel signals would cause a loss of information.
However, proving any of the above points of view would require more detailed modelling of changes in SNR when voxel signals are averaged to obtain the ROI time series.

As moving from the voxel level to the ROI level changes the link structure of the network, one may expect to see differences in other network properties as well. Indeed, \citet{hayasaka2010} have demonstrated that voxel and ROI-level functional brain networks differ from each other in terms of several network metrics such as degree distribution, characteristic path length, and local and global efficiency. These metrics naturally depend on the link structure of the network and may therefore have been affected by the structural changes caused by averaging voxel signals. Based on their results, \citet{hayasaka2010} decided to recommend against using anatomical ROIs as nodes of functional brain networks.

The voxels used in the acquisition and analysis of fMRI data are artificially defined and it is reasonable to assume that the true functional areas in the brain are larger than voxels \citep{wig2011, shen2013}. Therefore, coarse-graining of voxels into well-defined larger-scale regions can
yield more accurate results than voxel-level analyses. However, we have shown that all ROIs are not equally consistent. Therefore,
when working with ROIs, one should be careful in interpreting results and aware of possible problems, such as spuriously high correlations between inconsistent ROIs.

\section{Materials and Methods} \label{methods}

\subsection{Subjects}

13 healthy, right-handed subjects (11 females, 2 males, age 25.1 $\pm$ 3.9, mean $\pm$ std) participated in a study of emotion processing \citep{nummenmaa2014, saarimaki2016}. 
All subjects had normal or corrected-to-normal
vision, and none of them reported a history of neurological or psychiatric disease. All subjects were volunteers, 
gave written, informed consent according to the Declaration of Helsinki, and were compensated for their participation. The study was approved by the Research Ethics Committee of Aalto University.

\subsection{Data acquisition} \label{methods:acquisition}

Functional magnetic resonance imaging (fMRI) data were acquired with a 3T Siemens Magnetom Skyra scanner in the AMI Centre
(Aalto Neuroimaging, Aalto University, Espoo, Finland). A whole-brain T2*-weighted EPI sequence was measured with the 
following parameters: TR = 1.7s, 33 axial slices, TE = 24 ms, flip angle = 70 $^{\circ}$, voxel size = 3.1 x 3.1 x 4.0 mm$^3$, matrix size
64 x 64 x 33, FOV = 256 x 256 mm$^2$. Data from both an emotion-processing
task (reported in \citet{nummenmaa2014, saarimaki2016}) and an approximately 6 min (215 time points) resting-state session were recorded. 
Only the resting-state data were
used in the present study. During the resting state, subjects were asked to lay still with their eyes open, fixating to a gray background image,
and avoid falling asleep. 

Besides fMRI, anatomical MR images with isotropic 1 x 1 x 1 mm$^3$ voxel size were also acquired using a T1-weighted
MP-RAGE sequence.

\subsection{Preprocessing of the data} \label{methods:preprocessing}

The fMRI data were preprocessed with FSL \citep{smith2004, woolrich2009, jenkinson2012} and with an 
in-house MatLab toolbox, BraMiLa (\url{https://git.becs.aalto.fi/bml/bramila}). The preprocessing pipeline 
begun with removal of the 3 first frames of each subject's data in order to eliminate the error caused by scanner transient effect (leaving 212 time points for further analysis),
slice timing correction, motion correction by MCFLIRT \citep{jenkinson2002}, and extraction of white matter and 
cerebro-spinal fluid (CSF). Functional data were co-registered 
to the anatomical image with FLIRT (7 degrees of freedom), further registered to MNI152 2mm standard template (12 degrees of freedom), and resampled to voxels of  4 x 4 x 4 mm$^3$. 
Signals were linearly detrended, and signals from white matter and CSF were regressed out from the data. 

Regressing out from the data the global signal (GS) decreases motion-related variance of the data \citep{power2014}. However, removal of GS may also distort correlation patterns in
the network \citep{fox2009, gotts2013}. Therefore, there is no general consensus among the fMRI community if GS should be regressed out or not. In the present work, we decided not to regress out
GS. The potential effect of GS on our data would be increased ROI consistency (see below), since GS is shared among all voxels and therefore may increase
voxel-level correlations. However, we observed a wide range of ROI consistency values, including very low consistencies.

In order to control for motion artifacts, expansion of motion parameters was extracted 
out from the data with linear regression (36 Volterra expansion based signals, \cite{power2014}). Head motion has been reported as a source of artifacts in connectivity 
studies \citep{power2012}. Therefore,
we calculated the framewise displacement for each subject. However, as the framewise displacement of all the subjects was under the suggested threshold of 0.5 mm, we did not perform any
scrubbing. Further, in later analysis we also investigated wheter differences in head motion could explain differences in ROI consistency.

In order to further avoid artifacts, voxels that were located at the edge between brain and skull where the signal power was less than 2 \%
of the individual subject's mean signal power were excluded from further analysis. 

\subsection{Spatial smoothing} \label{methods:smoothing}

Our standard preprocessing pipeline did not include spatial smoothing. However, in order to investigate how this commonly used preprocessing method would affect our results,
we repeated all analysis with data that had been smoothed with a Gaussian kernel. We used 3 different kernel sizes: full width at half maximum (FWHM) of 5 mm, 8 mm, and 12 mm.
For details, see Supplementary Methods.

\subsection{Atlas-based Regions of Interest} \label{methods:rois}

After preprocessing (and spatial smoothing when applied), the cerebral cortex as well as subcortical structures and the cerebellum were divided into Regions of Interest (ROIs). For
defining the ROIs, we used three different brain parcellations: the commonly-used anatomical HarvardOxford (HO) and Automated Anatomical Labeling (AAL) atlases as well as the Brainnetome
atlas that is based on structural and functional connectivity.

The ROIs were defined so that each voxel belonged to one ROI only. The time series of ROIs were defined as 
averages of the time series of the voxels within the ROIs:

 \begin{equation}
  X_{I} = \frac{1}{N_{I}}\sum_{i\in I}x_{i}, \label{eq:roiseries}
 \end{equation}
where $I$ is the focal ROI, $N_I$ is its size defined as the number of constituent voxels, and $x_i$ is the time series of voxel $i$.

\subsubsection{HarvardOxford atlas}

HarvardOxford (HO) atlas (\url{http://neuro.debian.net/pkgs/fsl-harvard-oxford-atlases.html}, \cite{desikan2006automated}) is a probabilistic atlas that is based on the macroanatomical boundaries of 
the brain. In the present study, we used 138 ROIs from the HO atlas at a probability 
level of 30\% (meaning that each voxel belongs to the ROI it has been associated with in 30\% or more of the subjects in the group used
to create the parcellation). 96 of these ROIs were located at the cerebral cortex, 15 of them covered subcortical grey matter, and 27 were located in the cerebellum. Note that one
of the cerebellar ROIs of the HO atlas (ROI 120, Vermis Crus I) is not defined at the probability level of 30\% and was therefore excluded from the present study.

In the HO atlas, the distribution of ROI sizes is
broad: in the case of our in-house data, the number of voxels in a ROI varied between 5 and 857, median ROI size being 88 (for sizes of all ROIs, see Supplementary Table).

\subsubsection{Automated Anatomical Labeling atlas}

ROIs of the Automated Anatomical Labeling (AAL, \cite{tzourio2002automated}) atlas have been obtained by parcellating a spatially normalized high-resolution single-subject structural 
volume based on the main sulci. Then, a label has been automatically assigned to each of the ROIs. In the present study, we used 116 ROIs from the AAL parcellation. 90 of these ROIs
were located at the cerebral cortex, while 8 of them consisted of subcortical grey matter, and 18 covered the cerebellar cortex.

Similarly as in the HO atlas, also in AAL the size of ROIs varied across a wide range. In the case of our in-house data, the number of voxels in a ROI varied between 4 and 607, median
being 157 (for further details, see Supplementary Table).

\subsubsection{Brainnetome atlas}

The Brainnetome atlas \cite{fan2016human} is based on \textit{in vivo} structural and functional connectivity measured using multimodal neuroimaging techniques. In the present
study, we used 246 ROIs from the Brainnetome atlas, 210 of which were located at the cerebral cortex while 36 covered subcortical grey matter. Note that the Brainnetome atlas does not
include ROIs located in the cerebellum.

In Brainnetome, the size of ROIs varied less than in HO or AAL: in the case of our in-house data, minimum number of voxels in a ROI was 5 and maximum number of voxels in a ROI was 186, 
median being 63. So, Brainnetome ROIs were smaller than ROIs of HO or AAL.

\subsection{Network extraction} \label{methods:extraction}

We study functional brain networks at two different scales: at the level of voxels and at the level of ROIs. At the voxel level, nodes of the network represent single voxels of 
the cortical gray matter, whereas at the ROI level whole ROIs are used as
network nodes. At both levels, network extraction begins with calculating the weighted adjacency matrix $A$ between all
nodes so that its element $A_{i,j}$ indicates the strength of connection between nodes $i$ and $j$ of the
network. To define the connection strength, we calculate the Pearson correlation coefficients between time series of each $i$-$j$ pair. 
 Pearson correlation
is commonly used to quantify functional connections between nodes of brain networks (see \citep{wig2011, braun2012}). 
This procedure yields a full, symmetric,
weighted adjacency matrix that, after removing the diagonal values, 
contains $\frac{1}{2}N(N-1)$ independent real-valued elements where $N$ denotes the number of network nodes. 

We performed most of our analyses on the full adjacency matrix, using zero-lag correlations
between all nodes of the network. However, for the analysis of the relationship between consistency, degree, and 
strength of a ROI (see below), the network was thresholded to a set of densities (0.25, 0.5, 0.75, 1, 1.5, 
2, 2.5, 3, 3.5, 4, 4.5, 5, 10, 15, and 20\%). To obtain a network with a density $d$, links weaker
than the $1-d$th percentile were removed from the network by setting the corresponding elements in $A$ to zero. Thresholded networks were
analyzed as weighted, instead of transforming the thresholded adjacency matrix into a binary one. The range of densities was chosen to emphasize 
low densities, where network structure is most sensitive to small changes in link weights.

The neuroscientific interpretation of negative correlations between brain areas has been subject to dispute. It has been argued
that negative correlations are less reliable than positive, and should thus be excluded
from analysis \citep{shehzad2009, schwarz2011}. At the same time, others have argued that negative
correlations have a true neurobiological meaning \citep{wig2011, fox2009}. Negative correlations are indeed included in our analysis of full matrices, that is, calculations of distributions of correlations between time series of
voxels in same and different ROIs, and studies of the relationship between consistency and voxel and ROI-level correlations.
On the contrary, for thresholded networks that are used to investigate the relationship between consistency and network measures, the thresholding method automatically 
excludes negative correlations since only the strongest $d$\% of links
are accepted at each density $d$. 

\subsection{The consistency of ROIs} \label{methods:consistency}

In order to quantify the amount of variation between the time series of the voxels
within a ROI, we define the ROI consistency as

\begin{equation}
 \phi(I) = \frac{1}{N_I(N_I-1)}\sum_{i,i'\in I} C(x_i,x_{i'}),
 \end{equation}

where $I$ denotes the focal ROI, $N_I$ is the number of voxels in  $I$, $C(x_i,x_{i'})$ denotes the Pearson correlation coefficient between the time series $x_i$ and $x_{i'}$ of 
voxels $i$ and $i'$ that belong to $I$, and the summation is done over all voxel pairs within $I$. 
The interpretation of consistency is straightforward: if a ROI contains
many voxels with nonuniform dynamics, $\phi$ is low indicating that the ROI time series is a poor 
approximation of voxel dynamics. To the contrary, an ideal ROI where all voxels have identical time series would reach the theoretical
maximum value of $\phi = 1$. 

Negative values of consistency are possible only in theory. They arise only when the ROI consists of pairs of voxels with signals anticorrelated with each other and
independent of the signals of the rest of voxels in the ROI. In practice, there are only few significant negative correlations between signals of voxels within a ROI (see Fig~\ref{within_roi_correlations}),
and we did not observe any negative consistency values.

\subsection{Time-series correlations at the voxel and ROI levels}\label{methods:correlations}

In the present article, we ask whether consistency predicts the connectivity of a ROI. We do this  
by investigating
the relationship between the mean consistency and correlation between a ROI pair at two levels. 
The \emph{voxel-level correlation} between two ROIs
is defined as the mean correlation between the time series of voxels within the ROIs, averaged over all possible voxel pairs:

 \begin{equation}
 \left \langle C\left(I,J\right)\right \rangle_\mathrm{vox}= \frac{1}{N_{I}N_{J}}\sum_{i\in I}\sum_{j\in J}C(x_{i}, x_{j}). \label{eq:voxel-level}
 \end{equation} 

The \emph{ROI-level correlation} between two ROIs is  simply defined as the Pearson correlation between the time series of the ROIs:
\begin{equation}
C(I,J)_\mathrm{ROI} =  C(X_I,X_J), \label{eq:roi-level}
\end{equation}
where $X_I$ and $X_J$ are the ROI time series, computed as averages of voxel time series within $I$ and $J$ (Eq.~\ref{eq:roiseries}).

\subsection{Relationship between ROI consistency and network measures} \label{methods:nproperties}

In order to investigate the role of a ROI as a network node, we utilized two commonly used network measures: the degree $k$ and
the strength $s$ of a ROI. These measures are commonly used to identify the hubs of the functional brain network \citep{rubinov2010}.

The degree of a ROI is defined as the number of its neighbours, \emph{i.e.} the number of other ROIs to which it is 
directly connected by a link. The degree offers a rough estimate of how central a node is in the network: nodes with many neighbours can be
seen as more important for information flow than nodes with less neighbours.

The strength of a ROI is defined as the sum of the weights of the links of the ROI. Thus, it quantifies the amount of connectivity of the ROI. Like the degree,
the strength of a ROI can be used to estimate its centrality; unlike the degree, it accounts for link weights too. However, because strength mixes topology (existence of links)
with link weights, interpretation is not always straightforward: a ROI with a few strong links  can have the same strength as a ROI with many weaker links.

As discussed above, high ROI consistency is expected to be related to strong voxel-level correlations. 
It is also natural to assume that strong voxel-level correlations give rise to strong ROI-level correlations. Thereby, a natural hypothesis would be that ROIs that have high consistency are connected by strong links. As a result, these ROIs can be expected to have high strength, too. In a thresholded network, higher strength may also indicate larger number of links with above-threshold 
weight, giving rise to higher degree as well. We have investigated these relationships in networks thresholded to different densities.

\subsection{ABIDE data} \label{methods:abide}

In order to ensure that our results are not caused by any property specific to our in-house dataset, we repeated all our analysis for a second, independent dataset. This dataset, referred
to as the ABIDE data, was part of the Autism Brain Imaging Data Exchange I (ABIDE I) project \citep{di2014autism} and consisted of 28 healthy controls. For details of the ABIDE data, see
Supplementary Methods.

\section*{Acknowledgements}

We acknowledge the computational resources provided by the Aalto Science-IT project. We thank Rainer Kujala for inspiring discussions and for comments on the manuscript, and Marita Kattelus and Athanasios Gotsopoulos for 
their help in data acquisition.

\section*{Author contributions}

OK, MS, and JS designed the study. HS supplied data. OK and EG contributed tools for preprocessing and analysis. OK performed the analysis. OK and JS wrote the manuscript with the help
of comments of all authors.

\bibliographystyle{apalike}

\bibliography{references}

\pagebreak

\begin{center}
 \textbf{\large Supplementary Information for Consistency of Regions of Interest as nodes of fMRI functional brain networks}
\end{center}

\setcounter{equation}{0}
\setcounter{figure}{0}
\setcounter{page}{1}
\setcounter{section}{0}
\makeatletter
\renewcommand{\theequation}{S\arabic{equation}}
\renewcommand{\thefigure}{S\arabic{figure}}
\renewcommand{\bibnumfmt}[1]{[S#1]}
\renewcommand{\citenumfont}[1]{S#1}

\section{Supplementary Methods}

\subsection{ABIDE data}

In order to ensure that the results we have obtained are not caused by some special feature of our data, we repeated the analysis for a second, independent dataset. This dataset
was part of the data published by the Autism Brain Imaging Data Exchange I (ABIDE I) project \citep{di2014autism}. From now on, we will refer to this dataset as the ABIDE data.

\subsubsection*{Subjects}

The ABIDE dataset consists of data on 28 healthy subjects. These subjects were measured as part of the ABIDE I initiative at California Institute of Technology (Caltech; 19 subjects) and
at Carnegie Mellon University (CMU; 9 subjects). For analysis, subjects from both measurement sites were pooled into a single dataset.

The 19 subjects measured at Caltech had ABIDE subject IDs 51475, 51476, 51477, 51478, 51479, 51480, 51481,
51482, 51483, 51484, 51485, 51486, 51487, 51488, 51489, 51490, 51491, 51492, and 51493. 15 of the subjects were male, 4 female. Their age ranged between 17 and 56.2 years. 15 of 
these subjects were right-handed and one left-handed, while the handedness of 3 subjects
was ambiguous. None of the subjects had family history of autism spectrum disorders (ASD), and they reported no history of either ASD or other psychiatric or neurological disease. For
a detailed description of the Caltech data, see \cite{tyszka2014}.

The 9 subjects measured at CMU had subject IDs 50657, 50659, 50660, 50663, 50664, 50665, 60666, 60667, and 50668. All 9 subjects were male. 8 of the 
subjects were right-handed and one had ambiguous handedness. Their age ranged from 21 to 40 years. Subjects were healthy controls with no reported history of ASD or other psychiatric or
neurological condition.

\subsubsection*{Data acquisition}

For the subjects measured at Caltech, fMRI and MRI data were acquired with a 3 Tesla Magnetom Trio device (Siemens Medical Solutions, NJ, USA). Structural MR images were acquired with a
T1-weighted MP-RAGE sequence with $1\times1\times1 \text{mm}^3$ isotropic voxel size. The resting-state fMRI data were measured as a whole-head T2*-weighted EPI sequence with the following
parameters: TR = 2.0 s, TE = 30 ms, flip angle = 75 $^{\circ}$, voxel size = $3.5\times3.5\times3.5 \text{mm}^3$, matrix size = 64 x 64 x 34, FOV = 256 x 256 mm$^2$. The length of the data was 3 minutes
(150 time points). During the measurement, subjects were instructed to lie still with their eyes closed and prevent falling asleep.

At CMU, fMRI and MRI data were acquired with a 3 Tesla Magnetom Verio device (Siemens Medical Solutions, NJ, USA). Structural MR images were acquired with a T1-weighted MP-RAGE sequence
with isotropic voxel size of $1\times1\times1 \text{mm}^3$. The resting-state fMRI data were measured with a whole-head T2*-weighted EPI sequence with the following parameters:
TR = 2.0 s, TE = 30 ms, flip angle = 73 $^{\circ}$, voxel size = $3.0\times3.0\times3.0 \text{mm}^3$, matrix size = 64 x 64 x 20, FOV = 192 x 192 mm$^2$. The length of the resting state
data was 10.4 min (320 time points). During the measurement, subjects lay still and eyes closed in the scanner in a room with lights shut off.

\subsubsection*{Preprocessing and analysis}

The ABIDE dataset went through a preprocessing and analysis pipeline that was identical to the one applied on our in-house data (for details, see the main article). In order to avoid any
unpredictable effects caused by the different measurement parameters of the Caltech and CMU datasets, the anatomical images of different subjects were registered to MNI152 standard template
and resampled to voxel size of $4\times4\times4\text{mm}^3$ before creating the group-level masks for obtaining ROIs. Further, time series of different subjects were not averaged at any
point in the analysis. Instead, all averaging across subjects done at the level of correlations and consistencies.

Note that the parcellations used in the analysis of the ABIDE dataset slightly differed from those used for analysing the in-house data. In the case of the HarvardOxford (HO) atlas,
the ABIDE dataset contained only 119 anatomical ROIs, since ROIs 27, 28, 73, 74, 125, 126, 127, 128, 129, 130, 131, 132, 133, 134, 135, 136, 137, and 138 
(left and right inferior temporal gyrus, anterior division, left and right temporal fusiform cortex, anterior division, left, right, and vermis VIIb, left, right, and vermis VIIIa,
left, right, and vermis VIIIb, left, right, and vermis IX, left, right, and vermis X)
did not overlap across all the ABIDE subjects. 

In the case of the Automated Anatomical Labeling (AAL) atlas, the ABIDE data
contained 109 ROIs, since cerebellar ROIs 101, 102, 104, 105, 106, 107, and 108 (left and right cerebellum 7b, right cerebellum 8, left and right cerebellum 9, left and right cerebellum
10) did not overlap across ABIDE subjects. 

In the case of the Brainnetome atlas, the ABIDE dataset contained 241 ROIs, since ROIs 93, 94, 110, 118, and 125 
(left and rigth inferior temporal gyrus 7\_3, right parahippocampal gyrus 6\_1, right parahippocampal gyrus 6\_5, left superior parietal lobule 5\_1) did not overlap across subjects.
did not overlap across subjects.

\subsection{Spatial smoothing}

Spatial smoothing is a commonly used method in the preprocessing of fMRI data especially when the general linear model (GLM) is used as the analysis paradigm \citep{mikl2008}. 
Effects of spatial smoothing on the results of analysis of functional brain networks are not fully known. However, it is known that smoothing may affect network properties \citep{fornito2013}.
In order to investigate whether spatial smoothing changes our results, we applied spatial smoothing with a Gaussian kernel on our in-house dataset and repeated our analysis for this
spatially smoothed dataset using the HO atlas.

In the smoothing, the time series of each voxel was redefined as the
average of the time series of the voxel itself and its neighboring voxels, weighted by a Gaussian kernel. Smoothing was always applied before
coarse-graining the voxel space to ROI space, and before network extraction. The data were smoothed with
3 different kernel sizes: full width at half maximum (FWHM) of 5 mm, 8 mm, and 12 mm. These sizes are typically used among the fMRI community. 
It has been recommended in the literature that on one hand, the smoothing
kernel size ``should approximate the size of the underlying signal or evoked response'' that would be $\sim$3-5 mm on
the cortex \citep{hopfinger2000}. On the other hand, \cite{pajula2014} have recommended that the kernel size should be 2-3 times the voxel size, and
Mikl et al. (2008) have recommended a smoothing kernel with FWHM of approximately 8 mm or even larger for group-level studies. 

Effects of spatial smoothing on the network structure are sometimes compensated by discarding from the network links that connect nodes within the FWHM of the smoothing kernel 
from each other. However, these short-range links are strong also when smoothing has not been applied \citep{alexander2012, salvador2005}. Therefore, we decided to not remove 
these links to avoid unpredictable effects on the network structure.

\section{Supplementary Results}

\subsection{ABIDE data}

In order to ensure that our results are not explained by any features of our in-house dataset, we repeated all the analysis presented in the main article for the ABIDE dataset that is
independent from the in-house dataset. All essential results presented in the main article generalized also for the ABIDE dataset.

\begin{figure}
 \begin{center}
  \includegraphics[width=0.6\linewidth]{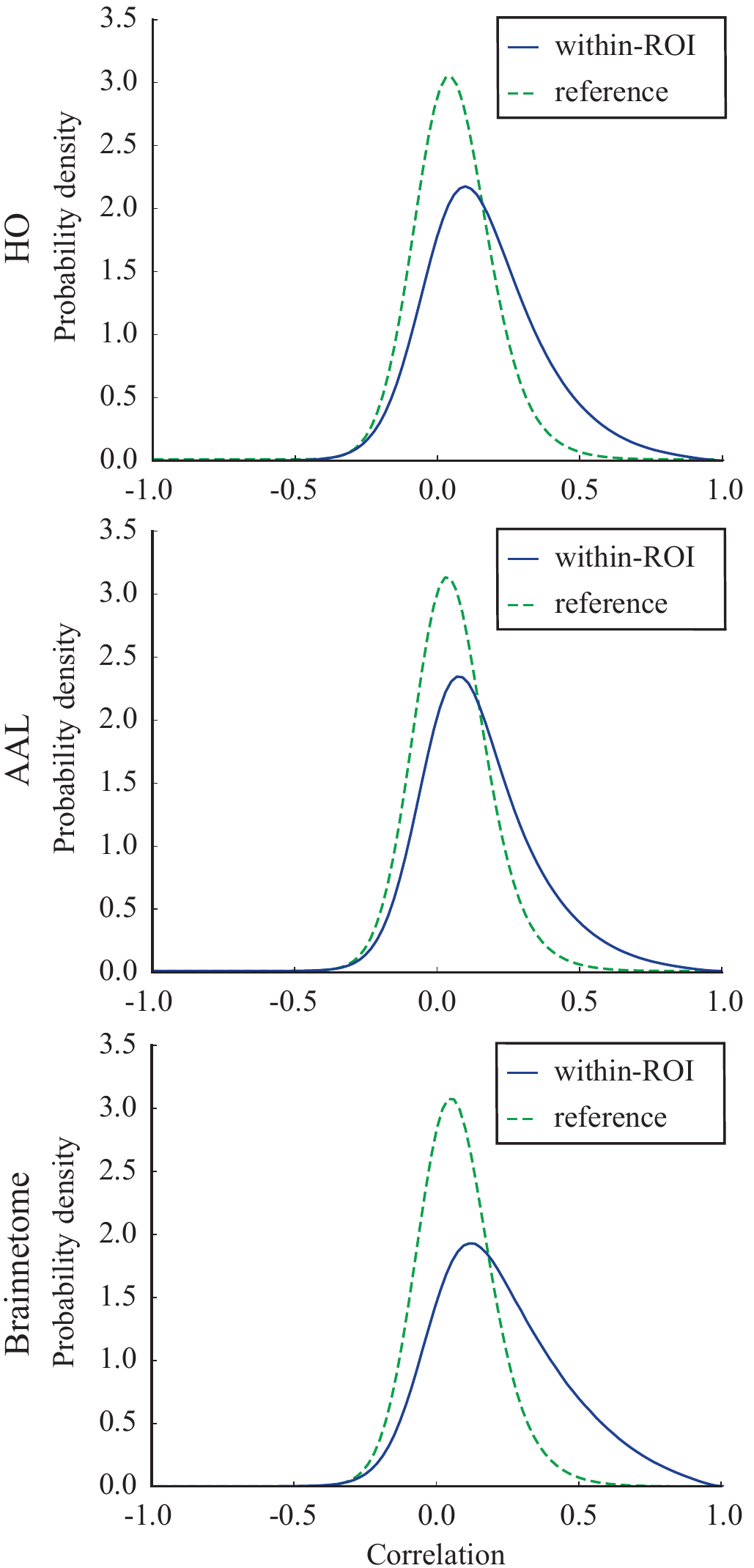}
  \caption{Correlation between voxels in the same ROI is on average stronger than voxels in different ROIs but not perfect. Distributions of Pearson correlation coefficients between voxel
  time series within (blue) and between (green, reference) ROIs have been calculated similarly as in Fig~\ref{within_roi_correlations} of the main article from the pooled data of 28
  ABIDE subjects. The large overlap of distributions indicates that ROIs are not functionally uniform.}
  \label{within_roi_correlations_ABIDE}
 \end{center}
\end{figure}

We started by calculating the distributions of Pearson correlation coefficients between time series of pairs of voxels that are in the same ROI and in different ROIs. In the Results
section of the main article,
we noticed that the distribution of in-ROI correlations is broad and largely overlaps with the reference distribution of correlations between voxels in different ROIs. This is true also
for the ABIDE data (Fig~\ref{within_roi_correlations_ABIDE}): although the in-ROI correlation distribution is more right-skewed and has a higher mean than the reference distribution
(HO: $r=0.15 \text{ vs } r=0.059$, Student's $t=37.64$, $p\ll10^{-5}$; AAL: $r=0.15$ vs $r=0.054$, $t=38.36$, $p\ll10^{-5}$; Brainnetome: $r=0.21$ vs $r=0.058$, $t=58.84$, $p\ll10^{-5}$)\footnote{The $p$-value has been calculated using a permutation-based two-tailed $t$-test, for details see \cite{glerean2016reorganization}.},
it also overlaps with the reference. This result indicates that in none of the three investigated atlases, ROIs are not functionally perfectly uniform neither in the ABIDE data.

\begin{figure}
 \begin{center}
  \includegraphics[width=0.8\linewidth]{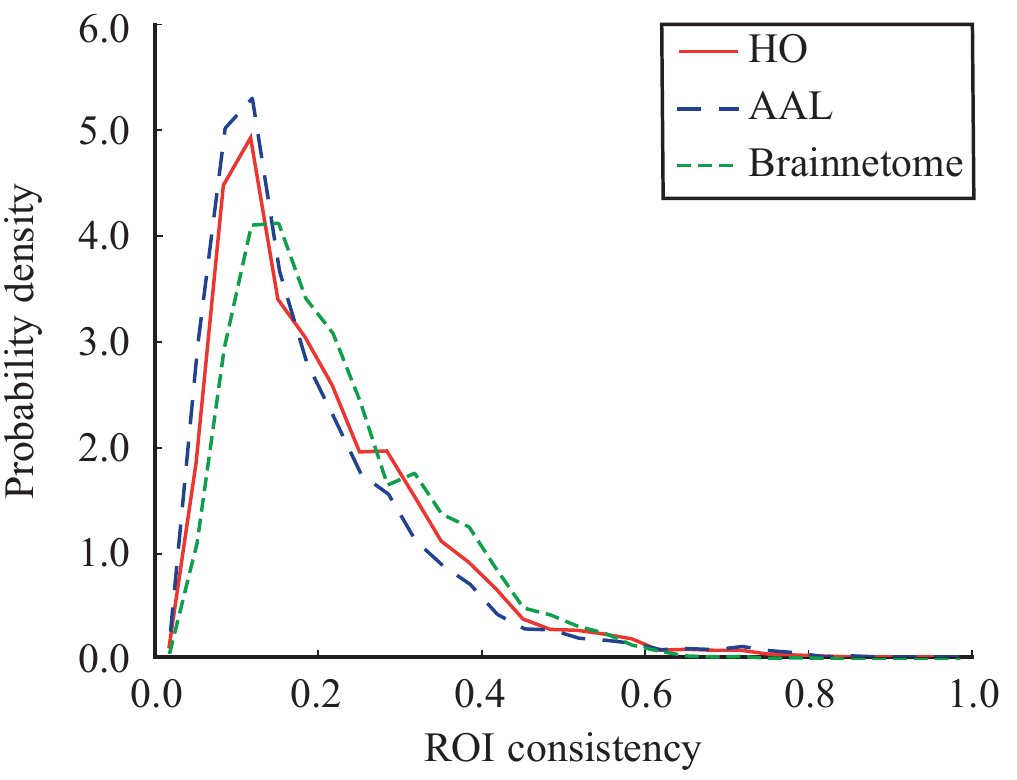}
  \caption{Consistency varies widely across ROIs in the ABIDE dataset. The consistency distributions has been calculated from the pooled data of 28 subjects.}
  \label{consistency_dist_ABIDE}
  \end{center}
\end{figure}

Next, we investigated the functional uniformity of the ROIs in terms of consistency (Fig~\ref{consistency_dist_ABIDE}). Similarly as in the case of the in-house dataset, the maximum value 
of consistency is high (HO: $\phi=0.80$; AAL: $\phi=0.86$; Brainnetome: $\phi=0.83$) but the distribution is broad and peaks at a relatively low consistency value (HO: $\phi=0.12$;
AAL: $\phi=0.12$; Brainnetome: $\phi=0.15$). 

\begin{figure}
 \begin{center}
  \includegraphics[width=0.8\linewidth]{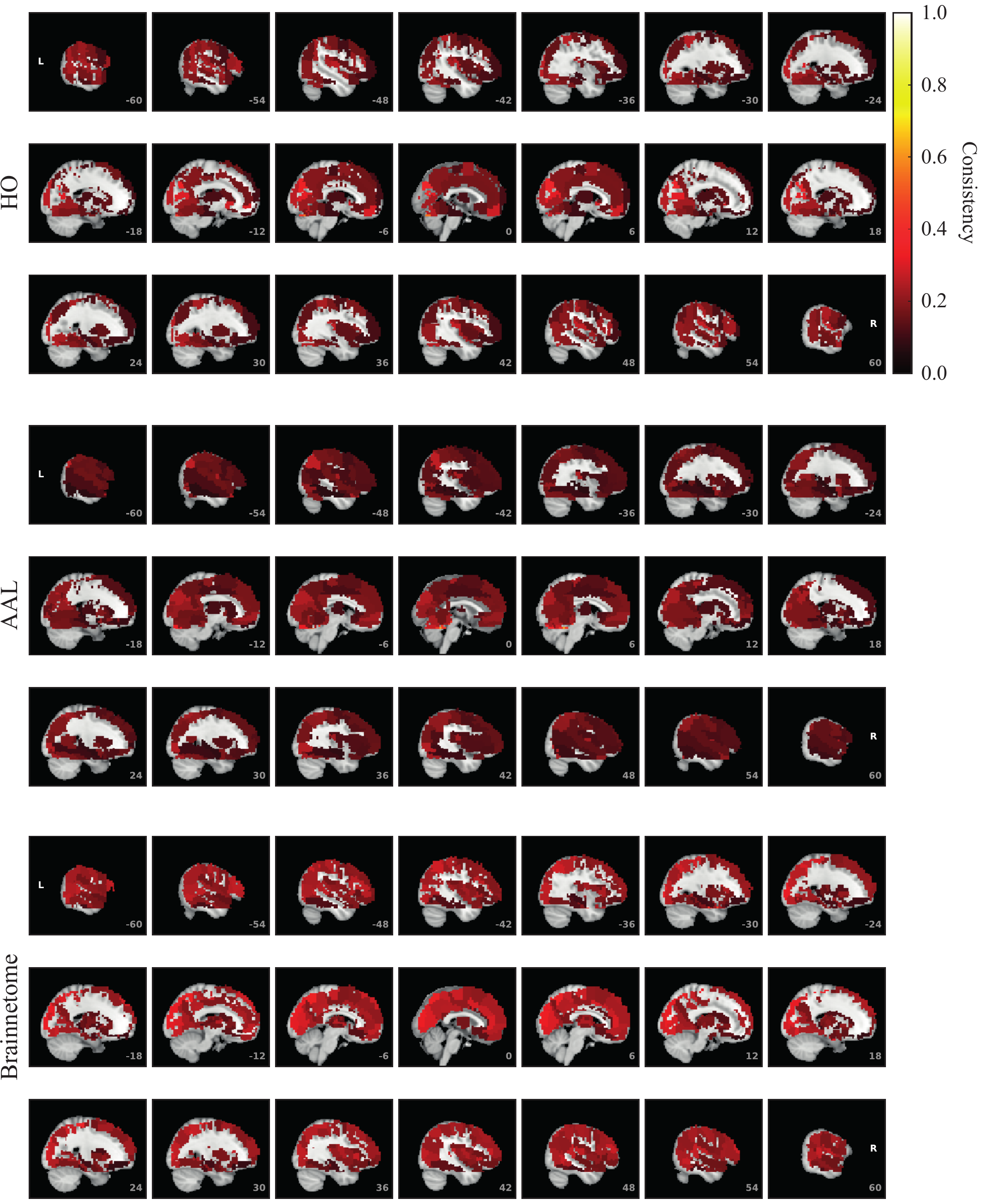}
  \caption{ROI consistency varies anatomically. We have visualized the mean consistency across 28 subjects 
  on a template brain similarly as in Fig~\ref{consistency_on_brain_surface} of the main article.} 
  \label{consistency_on_brain_surface_ABIDE}
 \end{center}
\end{figure}

From the visualization of the mean consistency across the 28 ABIDE subjects on a brain template (Fig~\ref{consistency_on_brain_surface_ABIDE}), we can see the wide variation of consistency:
the most consistent ROIs have twice the consistency of the least consistent ones. 

In HO, the most consistent ROIs included left and right supracalcarine and cuneal cortices, and cerebellar areas Left, Right, and Vermis Crus II. Importantly, these areas were among the
most consistent ROIs also in the in-house dataset.

In AAL, all the most consistent ROIs were small cerebellar ROIs: left and rigth Crus2, left cerebellar area 8, Vermis 1\_2, Vermis 8, Vermis 9, and Vermis 10. This set of most consistent
ROIs differs from the one obtained with AAL in the in-house dataset. This may indicate that the ROI homogeneity of the AAL atlas depends on the dataset used.

In Brainnetome, the most consistent ROIs included left precentral gyrus (6\_4), left parahippocampal gyrus (6\_1 and 6\_5), left superior parietal lobe (5\_1), and right postcentral
gyrus (4\_4). Similarly as in AAL, these ROIs were relatively small and not among the most consistent ones in the in-house data.

Similarly as in the in-house dataset, the intersubject variation was larger for
the least consistent ROIs than for the most consistent ones.
In HO, the least consistent ROIs included left precentral gyrus, right frontal pole, posterior division of left temporal fusiform cortex, left and right thalamus, brain-stem, and 
left and right hippocampus. In AAL, among the least consistent ROIs were left and right hippocampus, left parahippocampal gyrus, left temporal inferior gyrus, and left and right fusiform
cortex. In Brainnetome, the least consistent ROIs included left orbital gyrus (6\_3), left fusiform gyrus (3\_1 and 3\_3), and left and right hippocampus.
In all the atlases, the least consistent ROIs in the ABIDE dataset are partially
same as the least consistent ROIs in our in-house dataset. 

For sizes, mean consistency values, and consistency ranks of all ROIs in the ABIDE dataset, see the Supplementary Table.

\begin{figure}
 \begin{center}
  \includegraphics[width=0.6\linewidth]{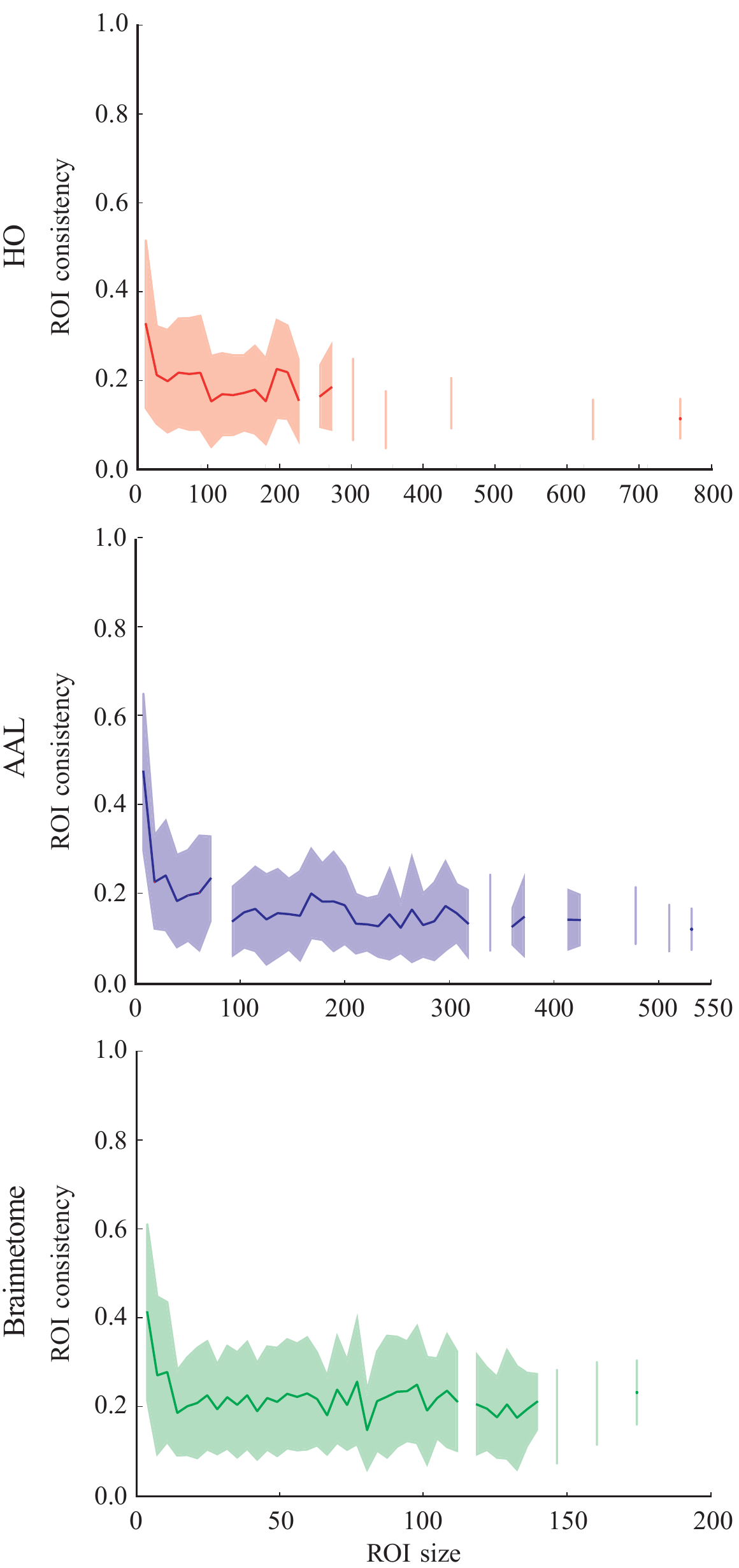}
  \caption{ROI consistency decreases with increasing ROI size in HO and AAL atlases. In the Brainnetome atlas, consistency does not depend on ROI size. The line shows the bin-averaged consistency and the shaded area shows the variation of consistency in each bin. Consistency
  has been pooled across 28 subjects, binned on the basis of ROI size, and bin-averaged, similarly as in Fig~\ref{roi_size_vs_consistency} of the main article.} 
  \label{roi_size_vs_consistency_ABIDE}
 \end{center}
\end{figure}

Similarly as in the in-house dataset, also in the ABIDE data ROI consistency is partially explained by its size in HO and AAL atlases (Fig~\ref{roi_size_vs_consistency_ABIDE}). Consistency is highest for small
ROIs and decreases with increasing ROI size (HO: Pearson correlation coefficient $r=-0.21, p\ll10^{-5}$; AAL: $r=-0.33$, $p\ll10^{-5}$). For the Brainnetome atlas, there is no correlation between ROI size and consistency 
($r=-0.070$, $p\ll10^{-5}$). Note that in the case of the Brainnetome atlas, the small $p$-value is explained by the large number of degrees of freedom (6720) and does not necessarily 
indicate significance.

\begin{figure}
 \begin{center}
  \includegraphics[width=0.8\linewidth]{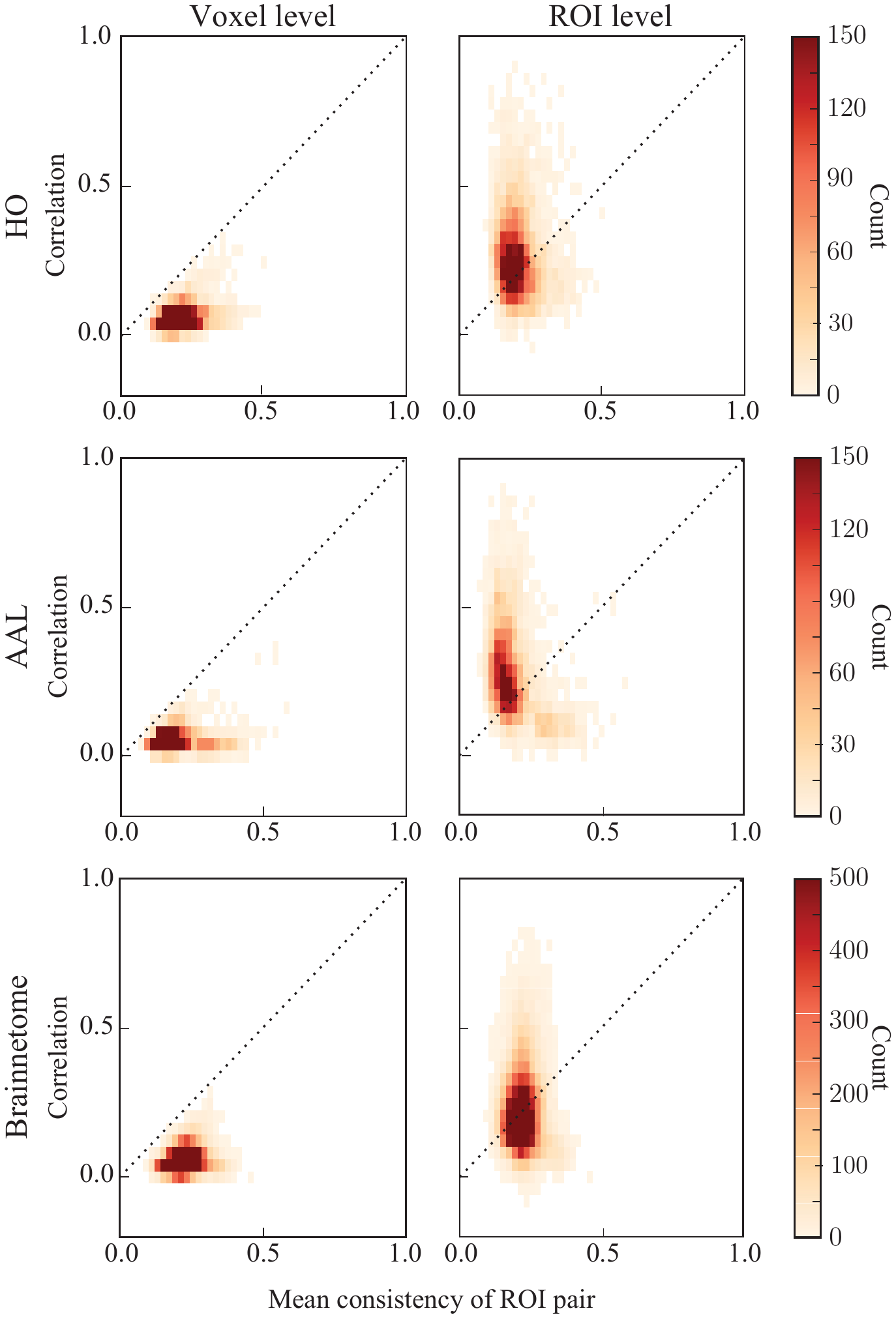}
  \caption{Left: High mean consistency is a prequisite for a strong voxel-level correlation between two ROIs. Right: ROI-level correlations are independent of mean consistency and also 
  low-consistency ROIs can be strongly correlated. In order to produce the heatmaps, the consistency and voxel and ROI-level correlations have been averaged across 28 subjects similarly
  as in Fig~\ref{consistency_vs_correlation} of the main article.}
  \label{consistency_vs_correlation_ABIDE}
 \end{center}
\end{figure}

In order to understand how a ROI's consistency affects its properties as a node of the brain network, we investigated how voxel and ROI level correlations between two ROIs behave as a
function of the mean consistency of this ROI pair. High mean consistency is a prequisite for high voxel-level correlation also in the ABIDE data, similarly as in in the case of our in-house
data. Although there is only a moderate correlation between ROI consistency and voxel-level correlation (HO: Pearson correlation coefficient $r=0.33, p\ll10^{-5}$; AAL: $r=0.11$, $p\ll10^{-5}$;
Brainnetome: $r=0.26$, $p\ll10^{-5}$), no strong voxel-level correlations
take place between low-consistency ROIs (Fig~\ref{consistency_vs_correlation_ABIDE}, left). ROI-level correlations, on the other hand, do not visibly depend on mean consistency
(Fig~\ref{consistency_vs_correlation_ABIDE}, right), although there is a weak negative correlation between consistency and ROI-level correlation (HO: Pearson correlation 
coefficient $r=-0.11$, $p\ll10^{-5}$; AAL: $r=-0.37$, $p\ll10^{-5}$; Brainnetome: $r=-0.015$, $p\ll10^{-5}$).

\begin{figure}
 \begin{center}
  \includegraphics[width=0.8\linewidth]{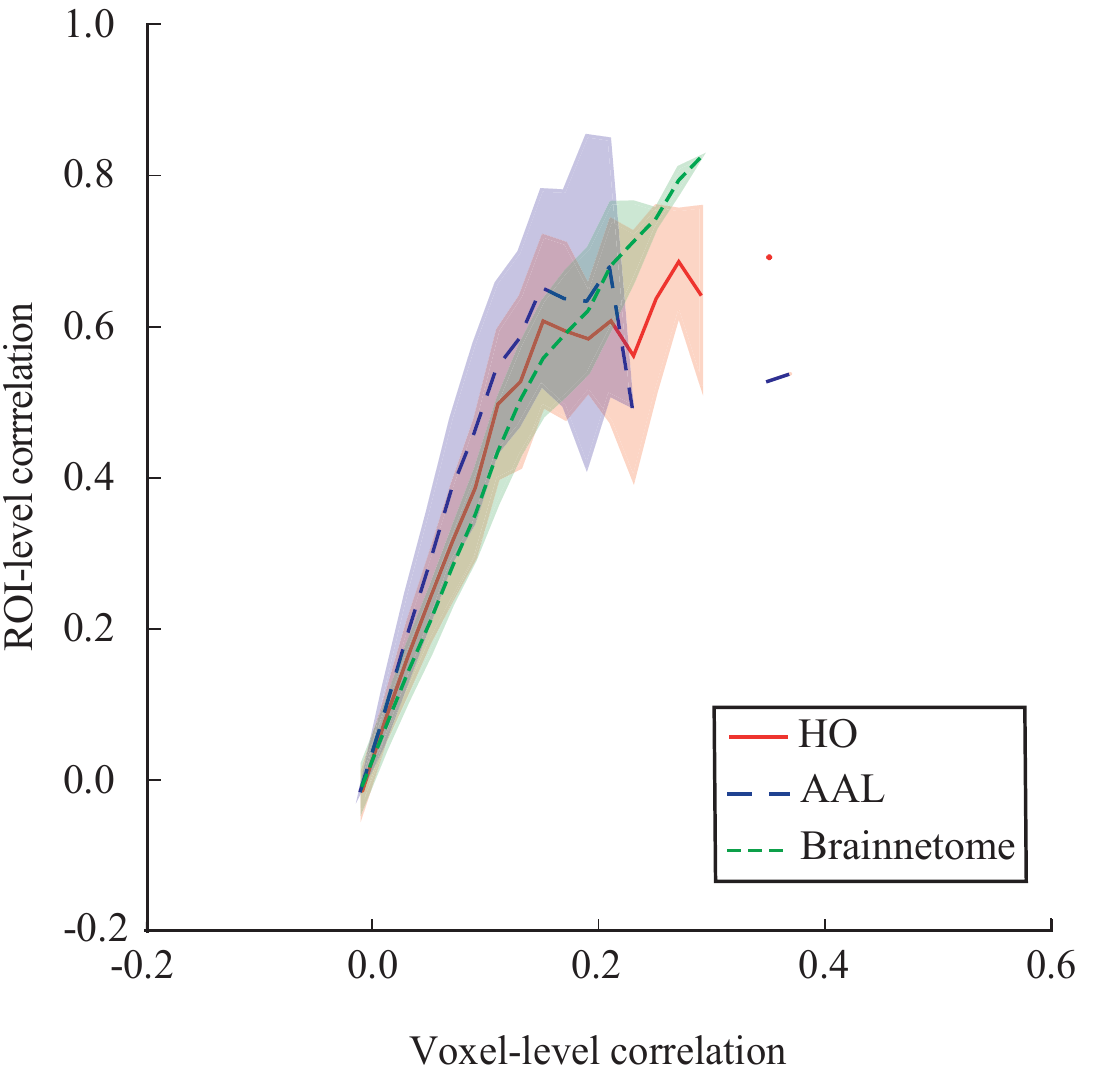}
  \caption{ROI-level correlations increase faster than the corresponding increasing voxel-level correlations. Averaging and binning as in Fig~\ref{roi_size_vs_consistency_ABIDE} and 
  Fig~\ref{roi_size_vs_consistency} of the main article.}
  \label{roi_vs_voxel_level_corr_ABIDE}
 \end{center}
\end{figure}

In the main article, we noticed that ROI-level correlations increase faster than the corresponding voxel level correlations. This is true also for the ABIDE dataset 
(Fig~\ref{roi_vs_voxel_level_corr_ABIDE}).

\begin{figure}
 \begin{center}
  \includegraphics[width=0.8\linewidth]{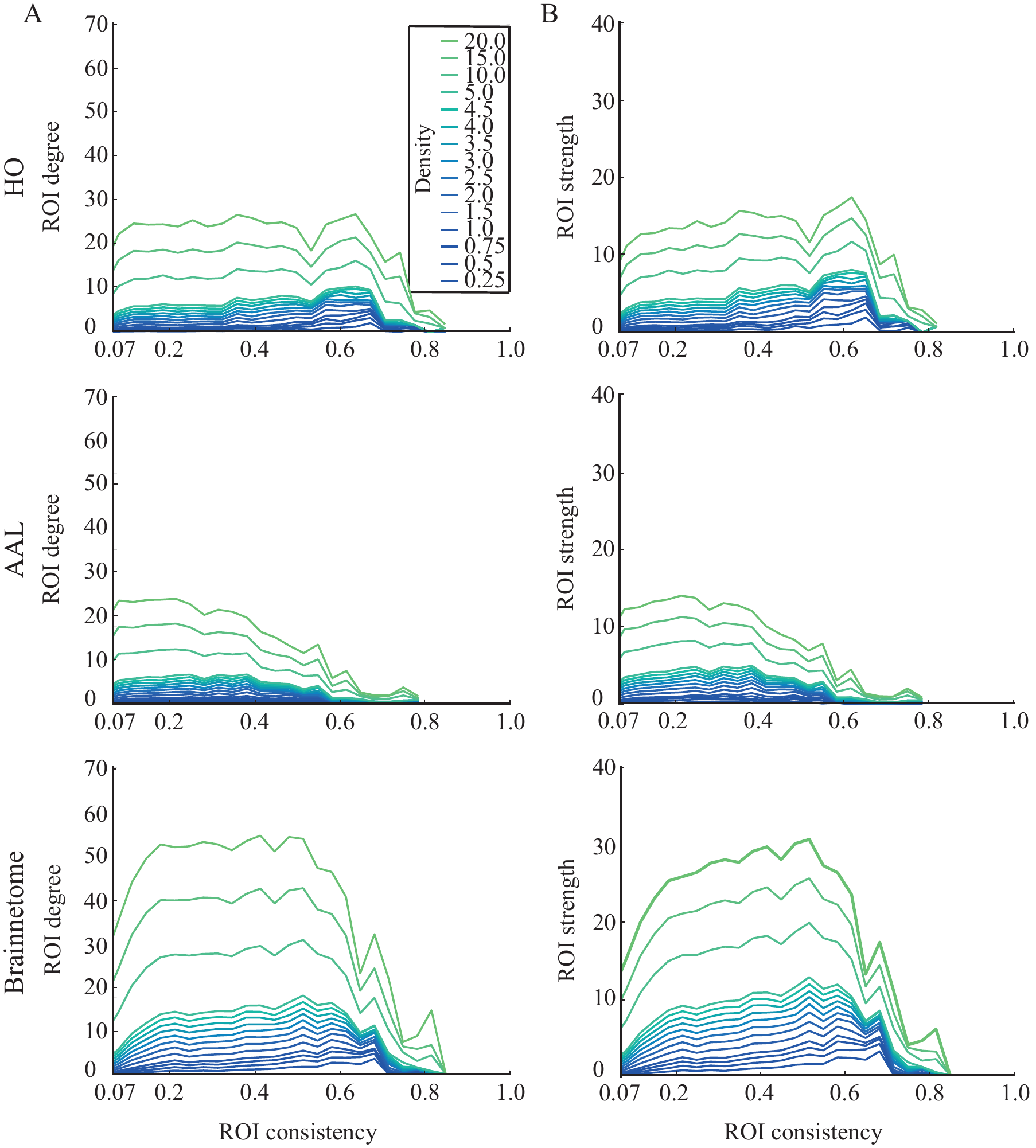}
  \caption{Degree (left) and strength (right) of ROI increase with the increasing consistency. Averaging and binning as in Fig~\ref{roi_size_vs_consistency_ABIDE} and 
  Fig~\ref{roi_size_vs_consistency} of the main article.}
  \label{consistency_vs_roi_properties_ABIDE}
 \end{center}
\end{figure}

In order to investigate in more detail the relationship between ROI consistency and it's network properties, we thresholded the network obtained from the ABIDE data to a set of densities $d$.
In these thresholded network, we calculated the mean degree and strength of ROIs across subjects. Both degree and strength increase with increasing consistency in HO and Brainnetome
(Fig~\ref{consistency_vs_roi_properties_ABIDE}). At $d=0.25\%$, for example, the Pearson correlation coefficient between consistency and degree was 
in HO $r=0.21, p\ll10^{-5}$, and in Brainnetome $r=0.28$, $p\ll10^{-5}$. At the same density, the correlation coefficient between consistency and strength was
in HO $r=0.22$, $p\ll10^{-5}$, and in Brainnetome $r=0.30$, p$\ll10^{-5}$. In AAL, instead, there is no correlation between neither consistency and degree ($r=0.089$, $p<10^{-5}$ nor
consistency and strength ($r=0.099$, $p<10^{-5}$) at $d=0.25\%$. Here, the small $p$-value is again explained by the large number of degrees of freedom (3052) and does not indicate significance.

In more dense networks, degree and strength depended less on consistency. At density $d=20\%$, no correlation was present between consistency
and degree in HO ($r=0.06, p=2,69*10^{-4}$), whereas between consistency and strength a correlation remained both at density $d=20\%$ ($r=0.16, p\ll10^{-5}$) and in the full network
($r=0.24$, $p\ll10^{-5}$).

In AAL, at $d=20\%$ there was a weak negative correlation between consistency and degree ($r=-0.16$, $p\ll10^{-5}$) and no correlation between
consistency and strength ($r=-0.078$, $p=1.8*10^{-5}$).

In Brainnetome, there was a correlation between both consistency and degree ($r=0.15$, $p\ll10^{-5}$) and consistency at strength ($r=0.25$, $p\ll10^{-5}$) at $d=20\%$. For consistency
and strength, this correlation remained also in the full network ($r=0.30$, $p\ll10^{-5}$).

\subsection{Spatial smoothing}

As mentioned above (Supplementary Methods), spatial smoothing may affect the results of functional brain network analysis. Therefore, we have
repeated all analysis of the main article for spatially smoothed data (for details, see Supplementary Methods).

\begin{figure}[]
  \begin{center}
      \includegraphics[width=0.8\linewidth]{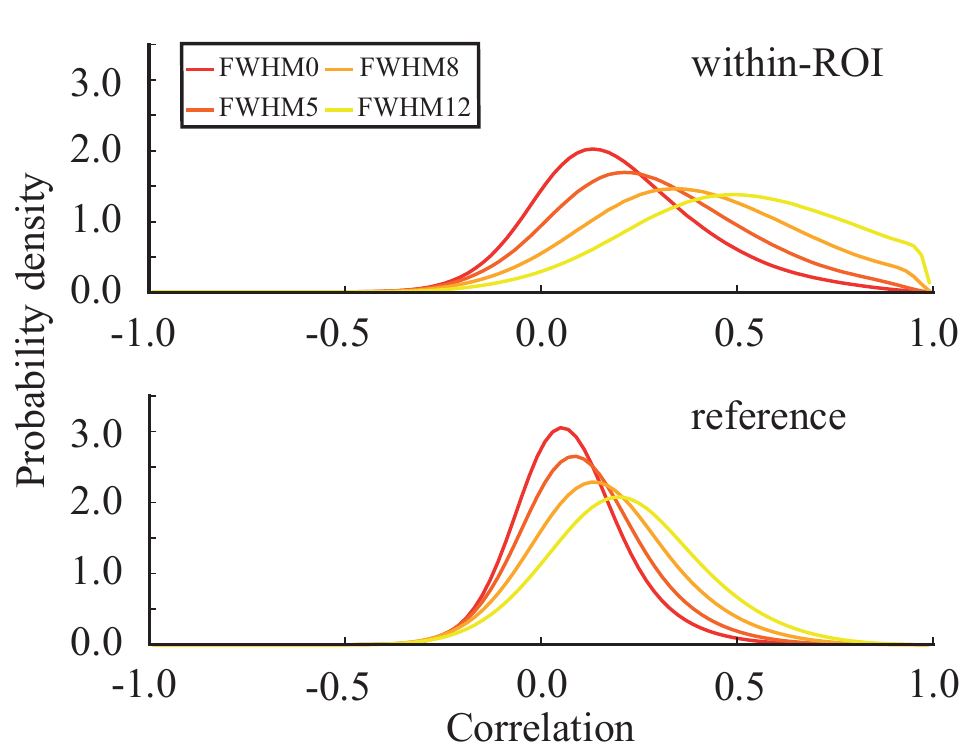} 
      \caption{Spatial smoothing increases correlations between voxel time series both within (upper part) and between (lower part, reference) ROIs. However, smoothing does not visibly 
      alter the left tails of the distributions. Correlation distributions are calculated across all ROIs from the pooled data of 13 subjects.}
      \label{within_roi_correlations_smooth}
 \end{center}
\end{figure}

We see that spatial smoothing increases correlations both within and between ROIs (Fig~\ref{within_roi_correlations_smooth}), shifting their distributions to the right
(distribution mean at FWHM0 $r=0.073$ vs $r=0.20$, Student's $t=49.81$, $p\ll 10^{-5}$; at FWHM5 $r=0.11$ vs $r=0.28$, $t=61.41$, $p\ll 10^{-5}$; at FWHM8 
$r=0.16$ vs $r=0.40$, $t=74.93$, $p\ll 10^{-5}$; at FWHM12 $r=0.22$ vs $r=0.51$, $t=89.42$, $p\ll 10^{-5}$)\footnote{$p$-value has been calculated using a permutation-based two-tailed $t$-test \citep{glerean2016reorganization}}. 
In the case of within-ROI correlations, smoothing alters the shape of the distribution, but does not remove the left tail even when using a kernel of full width at half maximum 
(FWHM) of 12 mm. Therefore, smoothing does not remove the functional inhomogeneity of the HarvardOxford ROIs. 

\begin{figure}[]
  \begin{center}
      \includegraphics[width=0.8\linewidth]{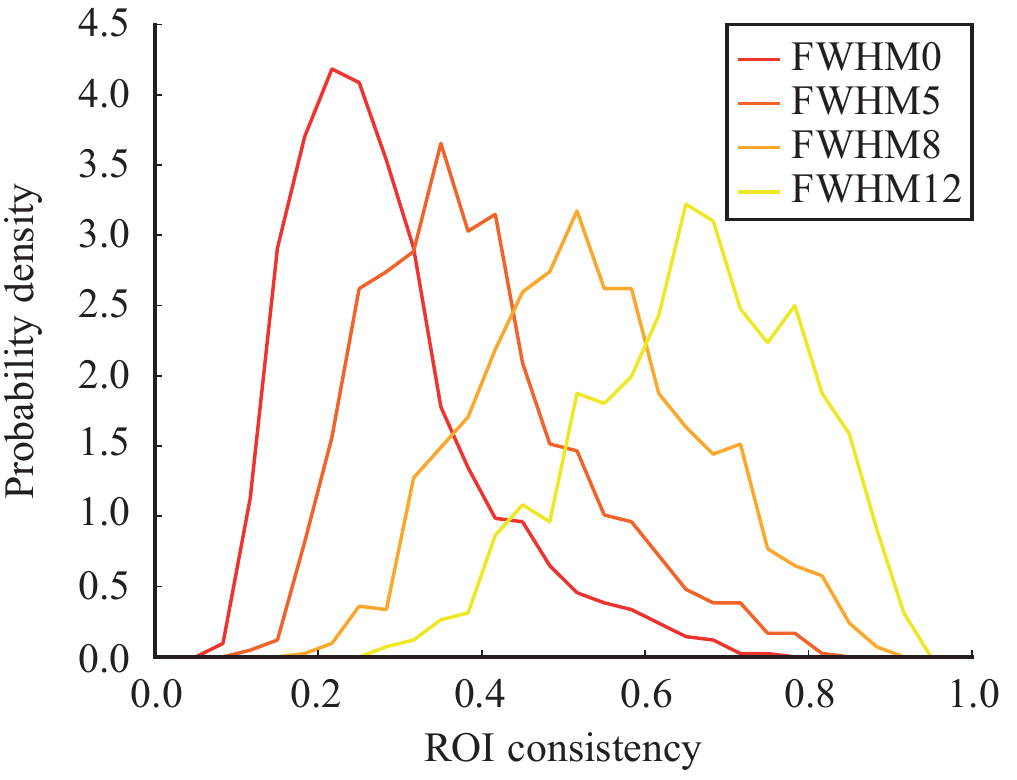}  
      \caption{Spatial smoothing shifts the ROI consistency distributions to the right but does not make them significantly more narrow. The consistency distributions have been calculated
from the pooled data of 13 subjects.}
      \label{consistency_dist_smooth}
 \end{center}
\end{figure}

\clearpage
Effects of spatial smoothing on the distributions of ROI consistency are similar as in the case of the correlation distributions.
We see that spatial smoothing generally increases consistency. However, smoothing only shifts the distribution to the right (Fig~\ref{consistency_dist_smooth}) without considerably narrowing it. 

\begin{figure}[]
  \begin{center}
      \includegraphics[width=0.8\linewidth]{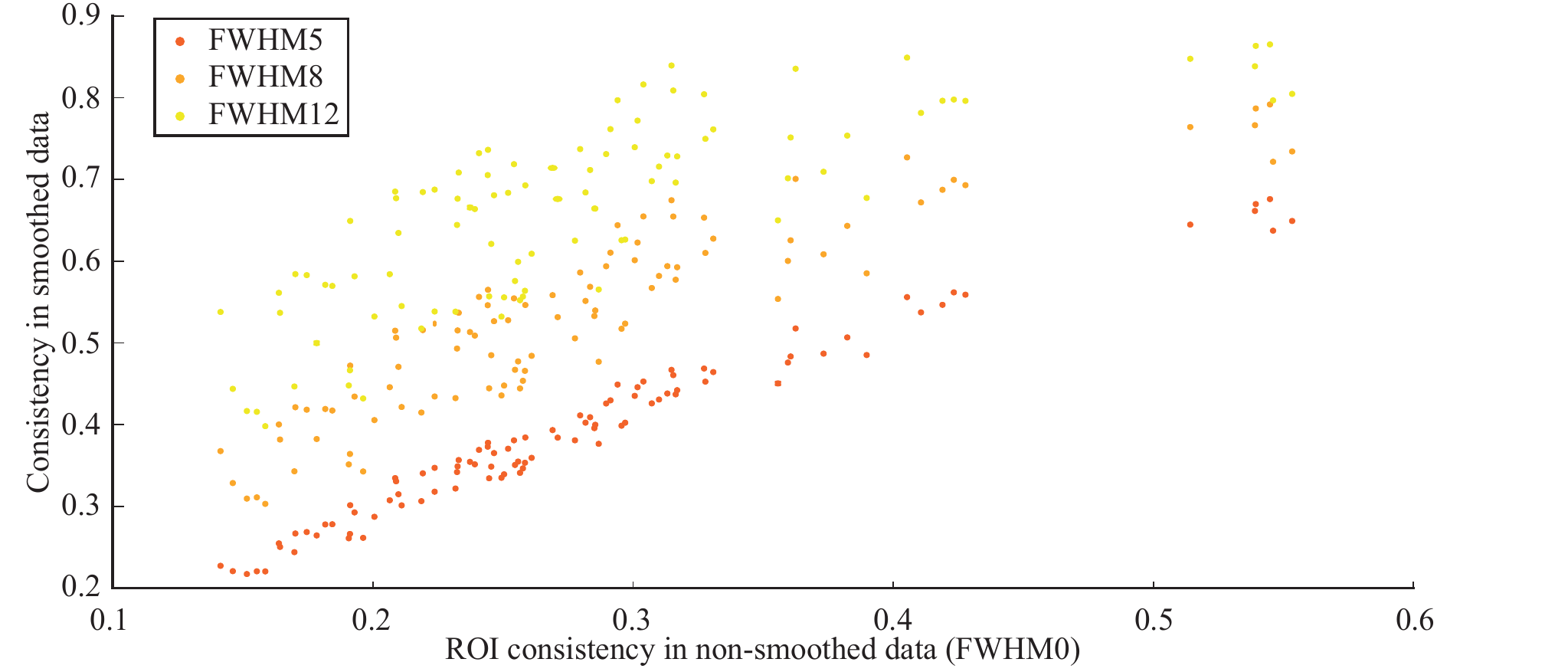}  
      \caption{Consistency in smoothed data correlates strongly with consistency in non-smoothed data. This indicates
 that the anatomical non-uniformity of consistency is not an effect of noise, nor is it eliminated by spatial smoothing.}
      \label{consistency_on_brain_surface_smooth}
 \end{center}
\end{figure}

Spatial smoothing does not change the 
identity of the most and least consistent ROIs: the same ROIs have the highest consistency both in the absence and presence of smoothing (Fig~\ref{consistency_on_brain_surface_smooth}; 
Pearson correlation between consistency in non-smoothed
data and data smoothed with FWHM5 $r=0.99, p\ll10^{-5}$, with FWHM8 $r=0.92, p\ll10^{-5}$, and with FWHM12 $r=0.81, p\ll10^{-5}$). 
This indicates that the anatomical variation of consistency
is not just an effect of noise that could be removed by filtering. 

\begin{figure}[]
  \begin{center}
      \includegraphics[width=0.8\linewidth]{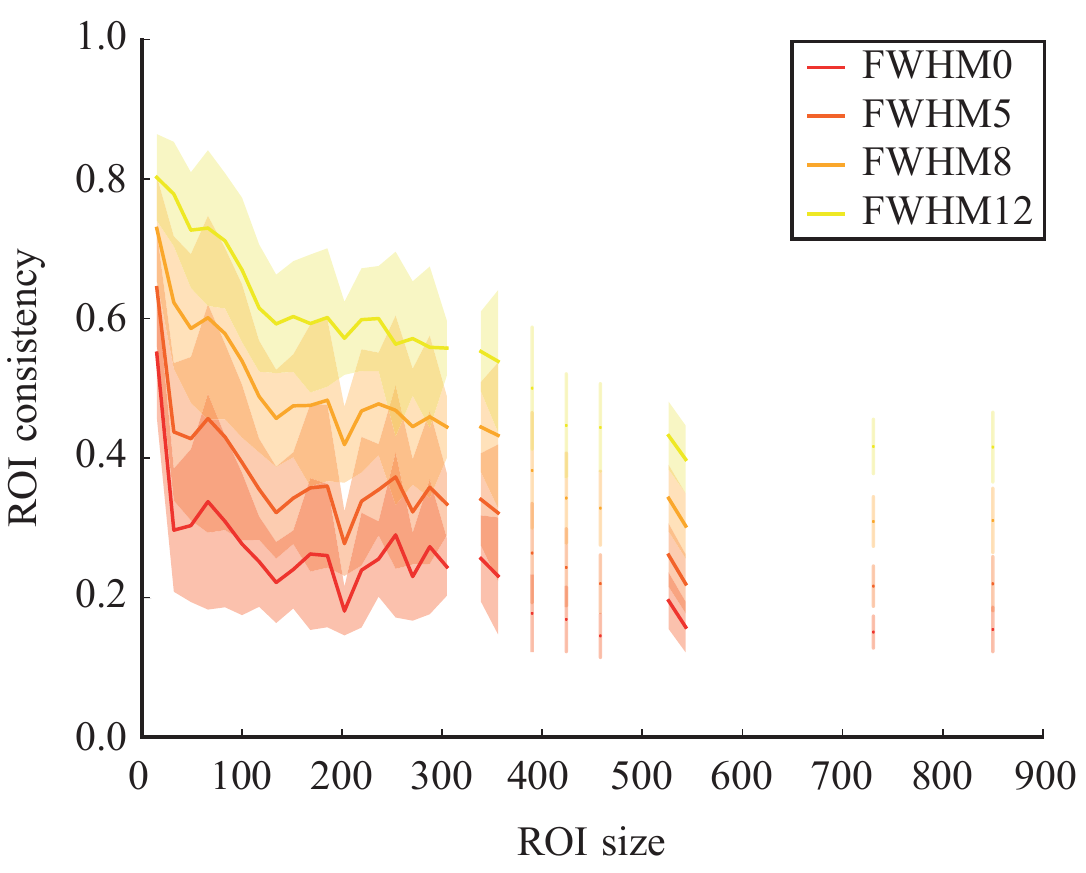}    
      \caption{Spatial smoothing does not affect how ROI consistency (lines) and its variation (shaded area) depend on ROI size. The consistency of each ROI has been pooled across
      13 subjects, binned on the basis of ROI size, and then bin-averaged.}
      \label{roi_size_vs_consistency_smooth}
 \end{center}
\end{figure}

The size of a ROI partially explains its consistency in the smoothed data too (Fig~\ref{roi_size_vs_consistency_smooth}). Negative correlation between the size and consistency of a ROI increases
with the increasing FWHM of the smoothing kernel (Pearson correlation coefficient $r=-0.35$ at FWHM0, $r=-0.43$ at FWHM5, $r=-0.54$ at FWHM8, $r=-0.64$ at FWHM12, $p\ll 10^{-5}$). 

\begin{figure}[]
  \begin{center}
      \includegraphics[width=0.8\linewidth]{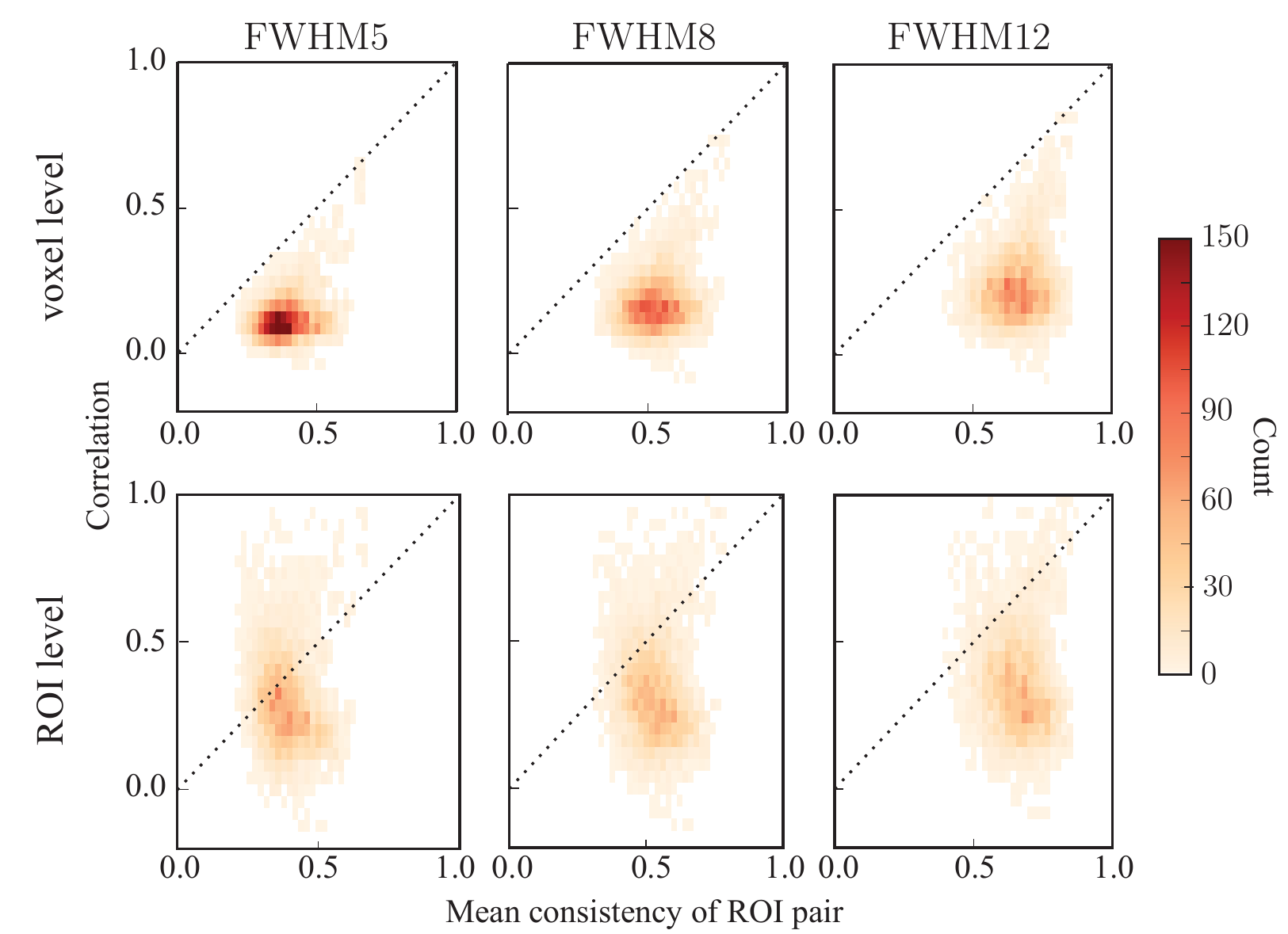}      
      \caption{Spatial smoothing does not affect the relationship between mean consistency and voxel and ROI-level correlations (upper and lower rows), 
      although consistency and voxel-level correlations are increased in presence of smoothing. In order to produce the heatmaps, the consistency and voxel and 
      ROI-level correlations have been averaged across 13 subjects.}
      \label{consistency_vs_correlation_smooth}
 \end{center}
\end{figure}

High consistency is required for high mean voxel-level correlation between two ROIs also when data have been smoothed, similarly to non-smoothed data (Fig~\ref{consistency_vs_correlation_smooth}, 
upper row). 
Spatial smoothing changes the Pearson correlation between consistency and voxel-level correlation slightly, and this change is typically negative 
(from $r=0.29$ at FWHM0 to $r=0.24$ at FWHM5, $r=0.17$ at FWHM8, and
$r=0.14$ at FWHM12, $p\ll 10^{-5}$). Spatial smoothing does not affect the relationship between consistency and ROI-level correlations neither (Fig~\ref{consistency_vs_correlation_smooth}, 
lower row; Pearson correlation $r=-0.14$ at FWHM0,
$r=-0.15$ at FWHM5, $r=-0.15$ at FHWM8, $r=-0.12$ at FWHM12, $p\ll 10^{-5}$).

\begin{figure}[]
  \begin{center}
      \includegraphics[width=0.8\linewidth]{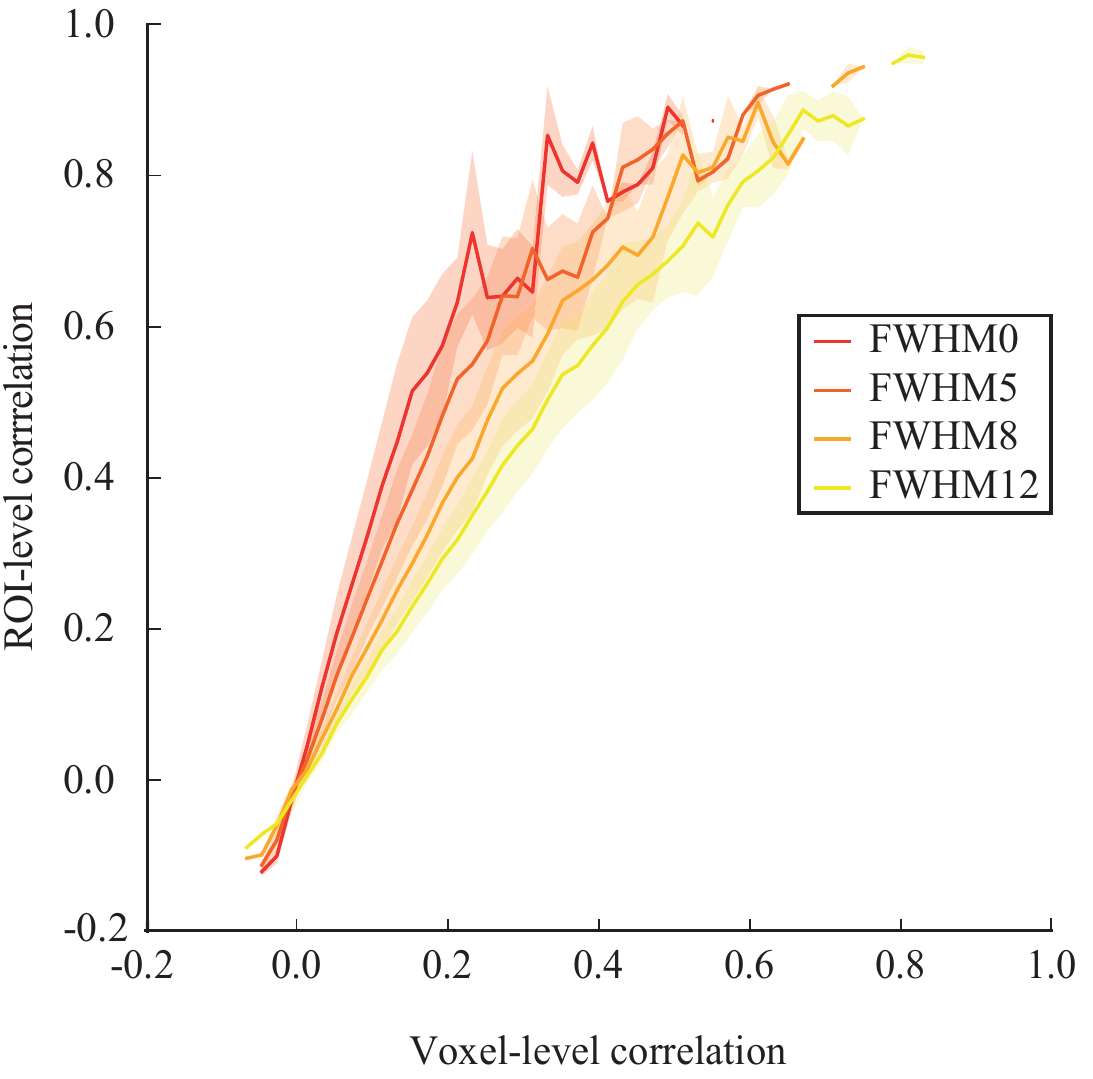}  
      \caption{Spatial smoothing decreases the difference between ROI and voxel level correlations. When smoothing is not applied, ROI-level correlations are stronger than voxel-level
      correlations because of the suppression of individual voxel-level signal components in when obtaining the ROI time series. However, smoothing suppresses individual signal components 
      already before the ROI time series are obtained. Voxel and ROI-level correlations have been averaged across 13 subjects and ROI-level correlations binned on the basis of voxel-level 
      correlations and bin-averaged.
      Averaging and binning as in Fig~\ref{roi_size_vs_consistency} of the main article.}
    \label{roi_vs_voxel_level_corr_smooth}
 \end{center}
\end{figure}

The nonlinear relationship between voxel and ROI-level correlations is more clearly visible without spatial smoothing (Fig~\ref{roi_vs_voxel_level_corr_smooth}). When smoothing is applied, 
individual voxel-level signal components get suppressed already before obtaining the ROI time series, and therefore moving to the ROI level has smaller effects.

\begin{figure}[]
  \begin{center}
      \includegraphics[width=0.8\linewidth]{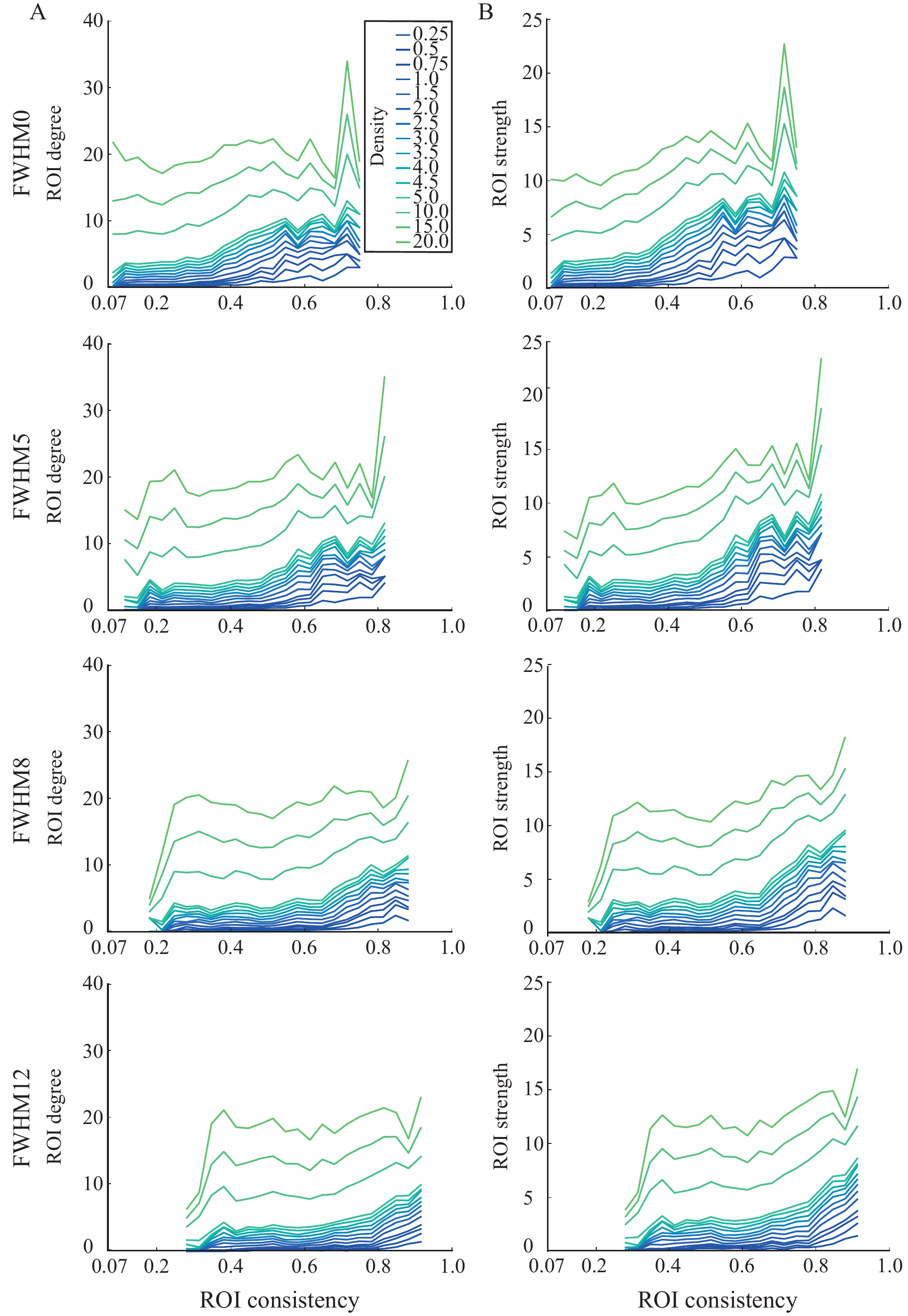}  
      \caption{The correlation between ROI degree (A) and strength (B) and consistency is visible independently of spatial smoothing. However, it is strongest when smoothing has 
      not been applied (uppermost row). Averaging and binning as in Fig~\ref{roi_size_vs_consistency} of the main article.}
    \label{consistency_vs_roi_properties_smooth}
 \end{center}
\end{figure}

\clearpage
The dependency between consistency, degree, and strength is weaker when spatial smoothing is applied (Fig~\ref{consistency_vs_roi_properties_smooth}; at $d=0.25\%$ and FWHM12, correlation for 
both degree and strength $r=0.27$, $p\ll 10^{-5}$). This is because smoothing suppresses individual voxel-level signal components similarly to averaging for obtaining ROI time series.

\section*{Supplementary Table}

The Supplementary Table is available at \url{https://github.com/onerva-korhonen/ROI_consistency} in Excel format (.xlsx). In the Supplementary Table, we present detailed information
about the ROIs used in the study. Besides the names and sizes of the ROIs, the Supplementary Table presents the consistency ranks of ROIs both in the in-house dataset and in the ABIDE
dataset.

\end{document}